\documentclass[a4paper,11pt]{article}
\usepackage{jcappub} 
\usepackage{orcidlink}
\usepackage{aas_macros}
\usepackage{amsmath}
\usepackage{graphicx}

\newcommand{\lya}{Ly$\alpha$}
\newcommand{\mpc}{$h^{-1}$Mpc} 

\title{Characterization of contaminants in the Lyman-alpha forest auto-correlation with DESI}

\affiliation{Affiliations are in Appendix \ref{sec:affiliations}}

\author[1]{{J.~Guy}\orcidlink{0000-0001-9822-6793},}
\author[1]{{S.~Gontcho A Gontcho}\orcidlink{0000-0003-3142-233X},}
\author[2]{{E.~Armengaud}\orcidlink{0000-0001-7600-5148},}
\author[3]{{A.~Brodzeller}\orcidlink{0000-0002-8934-0954},}
\author[4,5]{{A.~Cuceu}\orcidlink{0000-0002-2169-0595},}
\author[6,7,8]{{A.~Font-Ribera}\orcidlink{0000-0002-3033-7312},}
\author[9]{{H.~K.~Herrera-Alcantar}\orcidlink{0000-0002-9136-9609},}
\author[10,4,11,12]{{N.~G.~Kara{\c c}ayl{\i}}\orcidlink{0000-0001-7336-8912},}
\author[13]{{A.~Muñoz-Gutiérrez},}
\author[14]{{M.~M.~Pieri},}
\author[15]{{I.~P\'erez-R\`afols}\orcidlink{0000-0001-6979-0125},}
\author[7]{{C.~Ram\'irez-P\'erez},}
\author[16,2,17]{{C.~Ravoux}\orcidlink{0000-0002-3500-6635},}
\author[18,19]{{J.~Rich},}
\author[20,21]{{M.~Walther}\orcidlink{0000-0002-1748-3745},}
\author[2]{{M.~Abdul Karim}\orcidlink{0009-0000-7133-142X},}
\author[1]{{J.~Aguilar},}
\author[22]{{S.~Ahlen}\orcidlink{0000-0001-6098-7247},}
\author[23]{{A.~Bault}\orcidlink{0000-0002-9964-1005},}
\author[8]{{D.~Brooks},}
\author[1]{{T.~Claybaugh},}
\author[9]{{R.~de la Cruz}\orcidlink{0000-0001-9908-9129},}
\author[13]{{A.~de la Macorra}\orcidlink{0000-0002-1769-1640},}
\author[8]{{P.~Doel},}
\author[24,25]{{K.~Fanning}\orcidlink{0000-0003-2371-3356},}
\author[26,27]{{J.~E.~Forero-Romero}\orcidlink{0000-0002-2890-3725},}
\author[28,29,30]{{E.~Gaztañaga},}
\author[9,31]{{A.~X.~Gonzalez-Morales}\orcidlink{0000-0003-4089-6924},}
\author[32]{{G.~Gutierrez},}
\author[33]{{C.~Hahn}\orcidlink{0000-0003-1197-0902},}
\author[12,11,4]{{K.~Honscheid},}
\author[34]{{S.~Juneau},}
\author[35]{{R.~Kehoe},}
\author[23]{{D.~Kirkby}\orcidlink{0000-0002-8828-5463},}
\author[1]{{T.~Kisner}\orcidlink{0000-0003-3510-7134},}
\author[1]{{A.~Kremin}\orcidlink{0000-0001-6356-7424},}
\author[1]{{A.~Lambert},}
\author[1]{{M.~Landriau}\orcidlink{0000-0003-1838-8528},}
\author[36]{{L.~Le~Guillou}\orcidlink{0000-0001-7178-8868},}
\author[7,37]{{M.~Manera}\orcidlink{0000-0003-4962-8934},}
\author[12,4,10]{{P.~Martini}\orcidlink{0000-0002-4279-4182},}
\author[34]{{A.~Meisner}\orcidlink{0000-0002-1125-7384},}
\author[7,38]{{R.~Miquel},}
\author[39]{{P.~Montero-Camacho}\orcidlink{0000-0002-6998-6678},}
\author[40]{{J.~Moustakas}\orcidlink{0000-0002-2733-4559},}
\author[41]{{E.~Mueller},}
\author[42]{{A.~D.~Myers},}
\author[43]{{J.~Nie}\orcidlink{0000-0001-6590-8122},}
\author[44,9]{{G.~Niz}\orcidlink{0000-0002-1544-8946},}
\author[2,1]{{N.~Palanque-Delabrouille}\orcidlink{0000-0003-3188-784X},}
\author[45,46,47]{{W.~J.~Percival}\orcidlink{0000-0002-0644-5727},}
\author[1,48,49]{{C.~Poppett},}
\author[50]{{M.~Rezaie}\orcidlink{0000-0001-5589-7116},}
\author[51]{{G.~Rossi},}
\author[52]{{E.~Sanchez}\orcidlink{0000-0002-9646-8198},}
\author[1]{{D.~Schlegel},}
\author[53,54]{{M.~Schubnell},}
\author[55]{{H.~Seo}\orcidlink{0000-0002-6588-3508},}
\author[1]{{J.~Silber}\orcidlink{0000-0002-3461-0320},}
\author[34]{{D.~Sprayberry},}
\author[2]{{T.~Tan}\orcidlink{0000-0001-8289-1481},}
\author[54]{{G.~Tarl\'{e}}\orcidlink{0000-0003-1704-0781},}
\author[13]{{M.~Vargas-Maga\~na}\orcidlink{0000-0003-3841-1836},}
\author[43]{{H.~Zou}\orcidlink{0000-0002-6684-3997}}

\abstract{Baryon Acoustic Oscillations can be measured with sub-percent precision above redshift two with the Lyman-$\alpha$ (\lya) forest auto-correlation and its cross-correlation with quasar positions. This is one of the key goals of the Dark Energy Spectroscopic Instrument (DESI) which started its main survey in May 2021. We present in this paper a study of the contaminants to the \lya\ forest which are mainly caused by correlated signals introduced by the spectroscopic data processing pipeline as well as astrophysical contaminants due to foreground absorption in the intergalactic medium. Notably, an excess signal caused by the sky background subtraction noise is present in the \lya\ auto-correlation in the first line-of-sight separation bin. We use synthetic data to isolate this contribution, we also characterize the effect of spectro-photometric calibration noise, and propose a simple model to account for both effects in the analysis of the \lya\ forest. We then measure the auto-correlation of the quasar flux transmission fraction of low redshift quasars, where there is no \lya\ forest absorption but only its contaminants. We demonstrate that we can interpret the data with a two-component model: data processing noise and triply ionized Silicon and Carbon auto-correlations. This result can be used to improve the modeling of the \lya\ auto-correlation function measured with DESI.}

\begin{document}
\maketitle
\flushbottom

\section{Introduction}
\label{sec:Introduction}

Precisely measuring the expansion history of the universe allows one to discriminate between different theories of dark energy, the putative cause of the presently observed accelerated expansion of the universe \citep{Riess1998, Perlmutter1998,Frieman2008}.
One method to access the redshift-dependent expansion rate, and the probe of interest of this paper, uses baryon acoustic oscillations (BAO) \citep{Eisenstein2007}.
In this work, we investigate  pipeline-induced and astrophysical sources of systematic uncertainties in this measurement at high redshift.\\

The BAO-induced feature in cosmological correlation functions is used as a standard ruler whose characteristic scale has been measured with ever greater precision in the distribution of galaxies from large scale redshift surveys:
the Sloan Digital Sky Survey (SDSS) I and II \citep{Eisenstein2005, Percival2010, Ross2015},
the 2-degree Field Galaxy Redshift Survey (2dFGRS) \citep{Percival2001, Cole2005},
WiggleZ \citep{Blake2011}, the 6-degree Field Galaxy Survey (6dFGS) \citep{Beutler2011},
the Baryon Oscillation Spectroscopic Survey (BOSS) part of SDSS III \citep{Anderson2014}
and finally the extended BOSS survey (eBOSS) with \cite{Dawson2016, Alam2021} up to a redshift of $z < 2.0$. \\

Aside from galaxies, a powerful tracer of the large scale structure of the universe is the intergalactic medium.
Specifically, leveraging the series of absorption features of neutral hydrogen gas in quasars spectra, called the Lyman-$\alpha$ forest, has provided measurements of the BAO scale at redshift $z\,\sim\,2.3$ :
first with BOSS \citep{Busca2013, Slosar2013, Kirkby2013, Font_Ribera2014, Delubac2015, Bautista2017, dMdB2017}
and then with eBOSS \citep{dSA2019, Blomqvist2019, dMdB2020}.
Owing to the fact that ground-based large scale cosmological surveys observe in the optical range, the Lyman-$\alpha$ forest data collected allow one to probe the distribution of neutral hydrogen gas at $z\,>\,2.1$.
This redshift range is complementary to that of the galaxies observed by the same surveys.
At its best, the Lyman-$\alpha$ forest has provided an isotropic BAO measurement with a precision of 1.3\% \citep{dMdB2020}. \\

 The Dark Energy Spectroscopic Instrument (DESI) is the newest generation of fully operational ground-based multi-object spectroscopic instruments.
 Seeing first light on October 22, 2019 and starting the first of its five-year survey on May 14, 2021, DESI is expected to deliver percent-level BAO measurements from its $z\,>\,2.1$ quasar redshift sample and their \lya\, forests.
 This expected improvement in the precision of the \lya\, BAO measurement compared to previous surveys is due in part to the fact that DESI will observe quadruple as much $z\,>\,2.1$ quasars as that available in the final eBOSS sample.
 Another key point to delivering a strong cosmological measurement is a solid understanding of the systematics introduced by the DESI instrument and its pipeline, as well as the astrophysical systematics at our level of sensitivity.

 Studies of the sort have been conducted and mentioned in \cite{Bautista2017} and \cite{dMdB2020} to understand the systematics introduced by BOSS/eBOSS instruments and pipelines.
The dominant effect was found to be a small excess correlation in the \lya\ auto-correlation function for pixel pairs of the same wavelength. This effect results from a common noise introduced in neighboring fibers when subtracting the sky background.

We revisit those studies in the context of DESI. We first assess the spurious correlated signal introduced in the \lya\, auto-correlation by the spectroscopic pipeline (this does not affect the cross-correlation with quasars).
 We demonstrate our understanding of the source of those pipeline-induced correlations by modeling the sky background subtraction noise and the spectrophotometric calibration noise so they can be accounted for in the DESI analysis of the \lya\, autocorrelation \citep{RamirezPerez2023,Gordon2023}.
Then we  characterize the signal from astronomical contaminants due to foreground absorption.
To do so, we measure the auto-correlation of the quasar flux transmission fraction of low redshift quasars, where there is no \lya\, forest absorption but only its contaminants; i.e. other chemical species, or metals, such as triply ionized silicon (SiIV), triply ionized carbon (CIV) and ionized magnesium (MgII) \citep{Pieri2014letter}. In the same way that we define the \lya\ forest, the series of absorption features at wavelengths smaller than the SiIV (resp. CIV, resp. MgII) emission peak/doublet is called the SiIV forest (resp. CIV forest, resp. MgII forest). \\

 The paper is organized as follows.
 Section \ref{sec:data} presents the details of the DESI Survey and the first year dataset used in this study.
 Section \ref{sec:lya} gives an overview of the Lyman-$\alpha$ forest analysis. The data processing noise correlation is then characterized and modeled in section \ref{sec:systematics}. This model is validated when studying the auto-correlation of foreground absorbers in section~\ref{sec:astro}. Section \ref{sec:summary} provides a summary of our findings and the implications for the fit of the DESI \lya\ auto-correlation function.

\section{The DESI Survey}
\label{sec:data}

The Dark Energy Spectroscopic Instrument (DESI) project is measuring the redshifts of more than 40 million galaxies and quasars over 14,000 sq. deg. of the northern sky, effectively building the largest three dimensional map of the observable universe. Its goal is to measure the cosmic expansion history and the growth of large scale structure to great precision in order to gain insight into the nature of dark energy. \\

The DESI instrument is presented in detail in \cite{desi-instrument-overview-2022} and references therein; we provide here only a brief overview. DESI is a multi object spectroscopic system installed at the Mayall 4-m telescope at Kitt Peak National Observatory near Tucson in Arizona, USA. It features a prime focus instrument with a specially designed corrector and a focal plane comprised of 5000 robotic fiber positioners \cite{Silber2022} connected to ten 3-arm spectrographs. Specifically, the focal plane is divided into 10 \textit{petals} each containing 500 fibers. Each 500-fibers bundle is connected to one spectrograph. Each spectrograph is made up of a blue channel spanning 3600 \AA\, to 5930 \AA, a red channel spanning 5600 \AA\, to 7720 \AA\, and a near infrared channel spanning 7470 \AA\, to 9800 \AA. The spectrographs have a spectral resolution ranging from 2000 to 5000, see \cite{Guy2023}. A set of 5000 astronomical objects observed simultaneously by one pointing of the telescope is called a \textit{tile}.\\

The DESI spectroscopic pipeline is run immediately after the observations, once they have been uploaded from Kitt Peak down to the National Energy Research Scientific Computing Center, providing fully reduced spectra and redshifts within hours, along with quality diagnostics to validate the data set. The pipeline is also re-run with a uniform set of codes and calibrations for each data release.
In practice, this means: processing nightly calibration images (zeros, arcs, and flats), finding wavelength and line-spread-function solutions for each exposure, extracting the one-dimensional spectra from the two-dimensional frames, flat-fielding the spectra, subtracting a sky background model and calibrating fluxes, determining redshifts and classifications for each spectrum, and evaluating the quality of the data.
The spectroscopic pipeline is described in more detail in \cite{Guy2023}. Refer specifically to Figure 5 for a complete and detailed diagram of the spectroscopic pipeline data flow. \\

The survey, intended to operate for 5 years, is now well advanced. When the moon is below the horizon, also called dark time (see \cite{Schlafly2023}), the DESI survey observes three categories of astronomical objects of interest to be used as tracers of the matter density field. They are the Luminous Red Galaxies (LRG), the Emission Line Galaxies (ELG) and the quasars -- also referred to as Quasi Stellar Objects (QSOs). These tracers are spanning the range $0.4<z<4$, where the upper limit is defined by the scarcity of intrinsically bright high redshift quasars. Only quasars at redshifts $z > 2.1$ can be used for the study of \lya\ forests because of the UV cutoff of the spectrograph sensitivity and the atmospheric transmission. DESI targeted about 60 quasars per square degree (see \cite{Chaussidon2022}). At the end of the 5 year survey, we expect to have short of 1,000,000  \lya\, forests. That is $\sim4$ times more than BOSS and eBOSS. This will result in cosmological constraints from the \lya\, forest at the sub-percent level precision \citep{DESI_SV,HerreraAlcantar2024}. \\

To be able to exploit the information stored in the  \lya\, forests in the spectra of $z > 2.1$ quasars, the signal-to-noise requirements are higher than the fiducial requirements for redshift identification for all the other target classes. As such, quasars that will be used for \lya\, analysis need three to four times the amount of DESI fiducial dark time \citep{Schlafly2023}. \\

The state of each potential DESI target, i.e. whether they need extra observations, is tracked through the Merged Target List (MTL, see \cite{Myers2023}). Following each successful exposure and quality assurance check, the archived results of the spectroscopic analysis are used to update the MTL, adjusting the priorities of observed targets. Most importantly, newly detected \lya\, quasars are the highest priority targets in the main survey (before other galaxy classes) and should therefore be observed whenever possible.  \\

Apart from the signal-to-noise requirements in the \lya\, forest region, another key point is to accurately measure the redshift of the observed quasars.
To that end, the standard DESI pipeline (i.e. the automated spectroscopic data reduction), described in \cite{Guy2023}, applies a template-fitting code called \texttt{Redrock}~\citep{Bailey2024} to derive classifications and redshifts for each target.
This is combined with line-fitting “afterburner” codes, QuasarNET \citep{Busca2018,Farr2020} and a MgII afterburner, that are incorporated at the level of the MTL logic.
The MTL takes into account redshifts and redshift warnings from \texttt{Redrock}, as well as quasar classifications and redshifts from QuasarNET when updating the state of a given target. \\

The redshifts are used to update the MTL, promoting newly detected $z>2.1$ quasars to high priority targets which should be observed whenever possible.\\

The main survey was started on May 14, 2021. We use in this analysis spectra from the first year of data which will be published as part of the DESI first data release (DR1).
The quasar catalog used for this analysis relies on the DESI catalog from \cite{KP3}. Compared to the DESI Early Data Release \citep{Chaussidon2022}, this catalog is enriched with new redshifts calculated by using updated quasar templates  from \cite{Brodzeller2023} and an improved \texttt{Redrock} version. This addresses a bias on redshifts $z > 2$ identified through the \lya\, quasar cross-correlation \citep{Bault2024}. The updated redshift catalog is set for release alongside DR1. Following the definitions from Table 3 in \cite{Guy2023}, quasars passing a quality threshold\footnote{Quasars with ZWARN=0 or ZWARN=4, excluding the low $\Delta\chi^2$ flag. See definitions in \cite{DESICollaboration2023}.}
are retained, while those with reported pipeline issues are discarded.
Utilizing the HEALPix \citep{Healpix} coadded quasar spectra from the DR1 reduction of the main survey dark time program, we identify Damped Lyman-$\alpha$ Systems (DLAs) through two methods: Convolution Neural Network and Gaussian Processes (see \cite{Wang2022}). Notably, we refrain from rerunning the DLA finder with the new quasar redshift catalog due to the substantial time investment required. DLAs with neutral hydrogen column densities surpassing $2 \times 10^{20}$ cm$^{-2}$ are masked based on high-confidence detection and in accordance with the prescriptions from \cite{Wang2022}. Additionally, quasars with Broad Absorption Lines (BALs) are identified using the algorithm outlined in \cite{Filbert2023}, and contaminated regions in the spectra are masked following \cite{Ennesser2022} guidelines.\\

For the analysis presented in this paper, we use the MgII, CIV, SiIV and \lya\, forests. Depending on the forest considered, and therefore the corresponding redshift of the background quasar, we have short of half a million spectra at $z > 2.1$ and about a million at lower redshifts. The choice has been made to measure the transmitted flux fraction fluctuations in the observed wavelength range 3600 - 5772 \AA\, \citep{Gordon2023}, falling squarely onto the blue cameras of the DESI spectrographs. We define the following four distinct spectral regions of the quasar rest-frame spectrum: the \lya\, forest (1040 -- 1205 \AA), the CIV forest (1420 -- 1520 \AA), the SiIV forest (1260 -- 1375 \AA) and the MgII forest (1920 -- 2760 \AA).

\section{Analysis of the Lyman-$\alpha$ forest}
\label{sec:lya}

We provide a high level description of how we access cosmological information by looking at absorption features from quasar spectra.
More details can be found in \citep{Bautista2017,dMdB2020} for the BOSS and eBOSS analyses and \citep{RamirezPerez2023,Gordon2023} for the more recent analyses of the DESI Early Data.\\

\subsection{Correlation function estimator}
We use the \lya\ transmitted flux fraction $F(\lambda)$ in high redshift quasar spectra as tracer of the matter over density at redshift $z_\alpha=\lambda/\lambda_{\alpha}-1$, where $\lambda_{\alpha}=1215.67$ \AA\ is the Lyman-$\alpha$ transition wavelength in the Lyman series.
We call $\delta_q(\lambda)=F_q(\lambda)/\overline{F}(\lambda)-1$ the transmitted flux fraction fluctuation in a quasar spectrum $q$.
$\delta_q$ is obtained by dividing the observed quasar spectrum $f_q$ by an estimate of the unabsorbed quasar spectrum $C_q$  times the mean transmission flux fraction $\overline{F}$. $C_q$ is often referred to as \textit{continuum}.

\begin{equation}
  \delta_q(\lambda) \equiv \frac{f_q(\lambda)}{C_q \overline{F}(\lambda)} - 1
\label{eq:delta}
\end{equation}

The average flux transmission fraction $\overline{F}$ evolves smoothly with redshift following the variation of neutral hydrogen density and hence is a function of the observer frame wavelength $\lambda$. In practice we approximate the product $C_q \overline{F}(\lambda)$ by the average absorbed quasar spectrum times a correction to account for the specific amplitude and slope of the spectrum in the \lya\ forest:
$ C_q \overline{F}(\lambda) \simeq \overline{C F}(\lambda_{RF}) ( a_q + b_q \log \lambda )$, where $\lambda_{RF}$ stands for the quasar rest-frame wavelength. This is computed with an iterative procedure (as described in \cite{RamirezPerez2023}).\\

The estimator of the auto-correlation function that takes as input the transmitted flux fraction fluctuation is

\begin{equation}
\xi_A = \frac{\sum_{i,j\in A} w_iw_j \delta_i \delta_j }{\sum_{i,j\in A}  w_iw_j },
\label{eq:cf_estimator}
\end{equation}
where $i$ (resp. $j$) is an index that indicates a measurement on quasars $q_i$ (resp. $q_j$) at wavelength $\lambda_i$ (resp. $\lambda_j$) using the weights $w_i$ (resp. $w_j$) and the forest element $\delta_i$ (resp. $\delta_j$).
The weights are optimized to account for the intrinsic fluctuations introduced by cosmological large scale structures, as well as measurement noise.
Pairs (i, j) belonging to the same quasar are excluded to prevent contamination by correlated errors introduced by the estimation of $C_q(\lambda)$ for a given spectrum.
$A$ denotes the bin in the space of co-moving separation  $(r_{\perp},r_{\parallel})$ where $r_{\parallel}$ means distance of separation along the line of sight and $r_{\perp}$ refers to the distance of separation perpendicular to the line of sight.
We elect for the bins to be of 4 \mpc\, in both directions and evaluate the correlation function from 0 to 200 \mpc. \\

\subsection{Distortion due to the continuum fitting}

One should be aware that the measurement of the normalization and slope of a quasar spectrum in the computation of $\delta_q(\lambda)$ introduces correlations among the values of $\delta_q$ at different wavelengths in the same spectrum. To a good approximation, the resulting quantity is a linear combination of the original $\delta_q$. The linear transformation is actually enforced in the correlation function estimator where values $\hat{\delta}_{q}$ are computed from the $\delta_q$ by explicitly subtracting the weighted mean and slope along each line of sight. This linear transformation is called a ``projection'' in \cite{dMdB2020} (see their Eq. 5 and 6). It results that the measured correlation function of $\hat{\delta}_{q}$ in a separation bin $A$ is itself a linear combination of the correlation function of the original $\delta_{q}$ from different separation bins $B$. This set of linear coefficients defines the {\it distortion matrix} $D$ that transforms the auto-correlation of the $\delta_q$ field into that of the $\hat{\delta}_q$ field, $\hat{\xi}_A=D_{AB} \, \xi_B$.\\

\subsection{Contribution from other atomic transitions}\label{sec:lya-metal-mat}

The redshifts used to compute the comoving separations are based on the assumption that all absorption comes from the neutral hydrogen Lyman-$\alpha$ transition (hereafter at the wavelength $\lambda_\alpha$). In reality, other transitions from other elements, in particular silicon and carbon, also contribute to the measured transmitted flux fraction fluctuations. In consequence the absorption measured at a given wavelength can be understood as the combination of absorption occurring at different redshifts. An absorption observed at the wavelength $\lambda_i$ will receive a contribution from a Lyman-$\alpha$ absorption occuring at $z^{\alpha}_{i} \equiv \lambda_{i} / \lambda_\alpha -1$ but also contributions from other {\it metal} transitions $(m)$ from absorbers located at different redshifts $z^{m}_{i} \equiv \lambda_{i} / \lambda_m -1$. Let us now consider absorption observed at the wavelengths $\lambda_i$ and $\lambda_j$ and at angular separation $\theta_{ij}$ . We interpret this pair as a contribution to the measured Lyman-$\alpha$ correlation function assuming the longitudinal separation  $$r_{\parallel}(z^{\alpha}_i,z^{\alpha}_j) = \cos \left( \theta_{ij}/2 \right) \int_{z^{\alpha}_i}^{ z^{\alpha}_j} \frac{ c \, d z}{H(z)}$$
  but the same pair also receives contributions from the auto-correlation of the transition $m$ at $r_{\parallel}(z^{m}_i,z^{m}_j)$, and from the cross-correlation of the Lyman-alpha absorption and $m$ at $r_{\parallel}(z^{\alpha}_i,z^{m}_j)$ and $r_{\parallel}(z^{m}_i,z^{\alpha}_j)$, and similarly the contribution of the cross-correlation of $m$ with other transitions $m'$.

There is fortunately a limited set of other transitions that have a measurable contribution to the \lya\ forests. Considering each pair ($m,m'$) of possible transitions one at a time, one can compute, for each transition in the pair, the offset between the true and assumed redshift and compute a mapping matrix between the true comoving separation bins $B$ (using the redshifts $z^{m}_{i(j)}$ and $z^{m'}_{j(i)}$) and the assumed comoving separations bins $A$ (with the redshifts $z^{\alpha}_i$ and $z^{\alpha}_j$) of the correlation function.
This matrix $M$, called \textit{metal matrix} in previous works \citep{Bautista2017,dMdB2020}, is such that $\xi^{meas}_A = M_{AB} \xi^{true}_B$ where $\xi^{true}$ is the true correlation function of $m$ and $m'$, and $\xi^{meas}$ the corresponding contribution to the measured correlation function. The combination of those metal matrices can be used to interpret the measured correlation function.\\

We do not detail here the modeling of the \lya\, correlation function, and in particular how we use it to measure the BAO scale. Instead, we refer the reader to the model described in \cite{dMdB2020} and more recently in \cite{Gordon2023}.

\section{Correlated noise from the data processing}
\label{sec:systematics}

The study presented in this paper is concerned with sources of correlated noise that can contaminate the \lya\, auto-correlation function. The \lya\ auto-correlation function $\xi$ is obtained by correlating transmitted flux fraction fluctuations $\delta_q$ from different lines of sight (see Eq.~\ref{eq:cf_estimator}). It ensures that any source of noise that is not correlated from one quasar spectrum to the next will not bias the measurement. It follows that $\xi$ is robust to the large quasar spectral diversity that introduces large correlated residuals in the $\delta_q(\lambda)$ along a line of sight but no significant correlated signal from
one quasar to the next. Indeed, we do not expect the properties of quasars themselves to be correlated on large scales. In this section, we will address the sources of correlated noise introduced by the automated spectroscopic data reduction process that can contaminate the \lya\, auto-correlation function.
It is worth noting at this stage that in contrast, we do not expect any significant contamination in the \lya\ - quasar cross-correlation from the data processing noise. Indeed the quasar redshifts are primarily determined from the spectra at wavelengths larger that the \lya\ forest region, where broad emission lines are present. In consequence, we do not expect the redshifts to be systematically correlated with the noise realization in the \lya\ forest. We also do not expect that the target selection process results in a preferential selection of quasars with correlated spectral properties across the sky. This being said, one cannot categorically exclude the possibility of spurious signal coming from variations of imaging depth or calibration errors. The test performed in Section \ref{sec:astro} will address the robustness of the modeling carried out in this section.\\

The main steps of the spectroscopic pipeline, as described in \cite{Guy2023}, that could leave correlated residuals are the following:

\begin{itemize}
\item CCD image preprocessing: imperfect CCD bias and dark current subtraction and CCD gain fluctuations could leave correlated residuals for fibers with spectral traces on the same CCD amplifier.
\item Calibration of the spectrograph optics: incorrect wavelength calibration, fiber trace coordinates or Point Spread Function.
\item Spectral extraction: residual fiber to fiber cross-talk caused by bright sources leaving residuals in many adjacent fibers.
\item Correlated fiber flat fielding errors caused in particular by variations of the spectrograph throughput with humidity.
\item Correlated noise introduced in the subtraction of the sky background model (more details in \S\ref{sec:sky}).
\item Correlated noise introduced by the spectrophotometric errors (more details in \S\ref{sec:cal}).
\end{itemize}

Among those, the last two, namely the sky background model noise and the spectrophotometric calibration noise are known to contaminate the \lya\, auto-correlation function. This was first discussed in~\cite{Bautista2017}. We will characterize and model those contributions in the following sub-sections. The process followed will be the same in each sub-section. In order to match the survey characteristics of the \lya\, sample, we start from the same set of \{$\delta_q(\lambda)$\} used to calculate the \lya\, correlation (i.e. same coordinates, same redshifts, same weights) and replace the transmitted flux fraction fluctuation values by a realization of the residuals left by the sky subtraction process in \S\ref{sec:sky} and by the spectrophotometric calibration process in \S\ref{sec:cal}. We then model the correlated signal arising from the autocorrelation of these modified $\delta_q(\lambda)$.\\

In order to validate that the modeling done in this section reliably accounts for the spurious correlations introduced by the spectroscopic pipeline, we will conduct validation tests in Section \ref{sec:astro}.

\subsection{Correlated sky subtraction residuals}\label{sec:sky}
\subsubsection{Description}

To first order, the sky background that is subtracted from a science fiber spectrum is the weighted sum of the sky level measured in neighboring sky fibers at each wavelength. It results that the noise from this sky background is correlated for neighboring science fibers. Its subtraction leads to correlated science spectra, and consequently correlated transmitted flux fraction fluctuation $\delta_q$ in the \lya\ forest.  \\

The sky background subtraction is one important step of the spectroscopic data processing. The sky level is assuredly much larger than the signal from most DESI targets, and in particular from quasars. This imposes a stringent requirement on the accuracy of the sky background subtraction in order to minimize redshift errors. It was found in \cite{Guy2023} that the sky background continuum level was modeled at a precision of 1\% or better. Larger errors of 2 to 3\% were found on sky lines at wavelength larger than 6000 \AA. In the wavelength range of the DESI spectrographs used for Lyman-alpha studies (3600-5772 \AA, i.e. the blue channel), only the bright oxygen sky line at 5579 \AA\ is of concern.  In consequence it is masked out. The other lines being more than 10 times fainter (see \cite{RamirezPerez2023}) are not masked. One should note that even in the case where the sky background estimator is unbiased (thanks to a perfect modeling of the instrument response across fibers and wavelength among other things), the statistical noise from the fit of the sky background model (to a limited set of noisy fiber spectra) is still a source of concern. Indeed, once this noisy sky model is subtracted, it introduces a source of correlated noise among fibers. It is not the Poisson noise from the sky background in each fiber that is correlated among fibers, but the spectral residuals caused by the noise of the model that is subtracted to all fibers.\\

\begin{figure}[!]
\centering
  \includegraphics[width=0.75\linewidth]{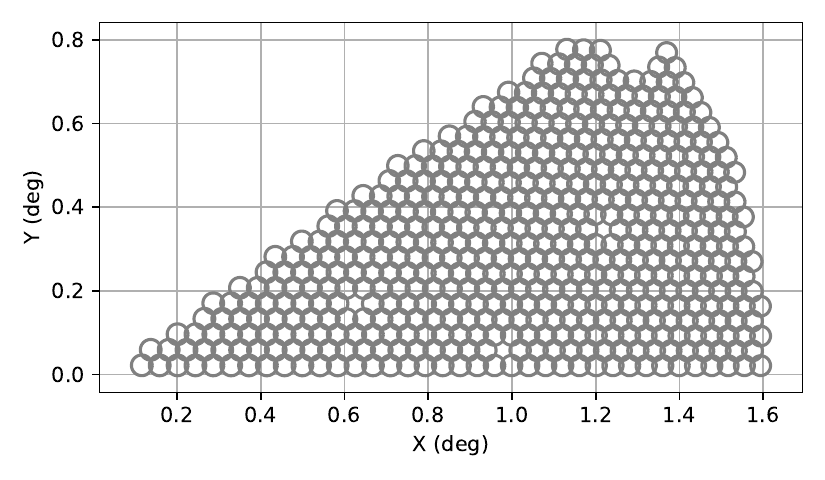}
  \caption{Patrol areas of the 500 fiber positioners of a DESI petal with physical coordinates converted to angles to the center of the field of view. One can see that the patrol areas from adjacent fibers overlap.\label{fig:desi-petal}}
\end{figure}

The sky background modeling and its subtraction is performed independently for each exposure and for each petal of the focal plane. It is based on a fit of the spectra from a dedicated set of fibers purposefully pointed towards blank sky. We call them sky fibers. For each petal of 500 fibers, a minimum number of 40 sky fibers are allocated for each observation. In practice there are often more sky fibers because the fibers of broken positioners are also included in the fit when they are not pointing to sources. More information about the choice of blank sky locations can be found in \cite{Myers2023}. The angular length of a DESI petal is 1.6 degree (see Figure~\ref{fig:desi-petal}). This corresponds to a comoving transverse separation $r_\perp$ of roughly 110~\mpc\ at redshift 2.4 for a fiducial $\Lambda$CDM cosmology with $\Omega_M = 0.315$. We do not expect to find a correlated signal from the sky subtraction at larger separations.

In \S\ref{sec:sky-noise-characterization}, we characterize in more detail the contribution from correlated sky subtraction residuals. In \S\ref{sec:sky-model}, we give the accurate form of this contribution as a function of $r_\perp$.

\subsubsection{Characterization}
\label{sec:sky-noise-characterization}
We intend to evaluate the contribution of the sky model noise to the \lya\ auto-correlation of the DESI Y1 sample. As stated in the introduction of \S\ref{sec:systematics}, we start from the set of forest $\{\delta_q\}$ obtained for this sample. This data set consists of lines of sight from more than 450,000 quasars, the details of which are published with the Year 1 analysis \citep{KP6}. We note that the procedure used to extract the $\delta_q$ from the spectra is the one presented in \cite{RamirezPerez2023}.\\

We build a synthetic data set by replacing the original $\delta_q$ values of transmitted flux fraction fluctuations by a realization of the sky subtraction residuals while keeping exactly the same weights and sky coordinates. This guarantees that the mock sample will be directly comparable to the true data, in particular in terms of redshift distribution.\\

The original $\delta_q$ vector of each quasar has been calculated starting with the combined spectrum obtained by optimally averaging the spectra of the same quasar from several exposures (up to four exposures to reach full depth in the DESI main survey). As part of the DESI pipeline processing, a sky background spectrum model is derived for each exposure and each of the 30 cameras (considering 3 cameras per DESI spectrograph). Our focus is on the blue cameras spanning the wavelength range 3600–5772 \AA, which fully contain the \lya\ DESI Y1 sample. The sky background model is composed of a common spectrum applied to all the fibers of a camera, along with corrections on sky lines that are based on templates obtained with a principal component analysis (see \cite{Guy2023}, Eq. 15). When it comes to the wavelength range used for the \lya\ forest, we can ignore those corrections as they affect only the bright oxygen sky line that is masked out.\\

Consequently, in order to emulate the spectral correlations coming from the sky model noise in each blue camera and exposure, we can simply generate a random realization of the common sky spectrum noise assuming that it follows a Gaussian distribution with zero mean and with the same variance as the one provided by the pipeline for the sky model (which has been modeled in great details, see Equation 16 of \cite{Guy2023}), all the while being uncorrelated from one wavelength bin to the next. We apply to the noise vectors the flux calibration derived from standard stars for the same exposure and camera, such that the resulting $\delta F_{sky}(\lambda)$ have the same units as the calibrated quasar spectra. One realization of $\delta F_{sky} (\lambda)$ applies to all of the fibers from the same camera and the same exposure.\\

Then, for each \lya\ line of sight, we identify the list of exposures and cameras that were used to produce the combined sky-subtracted spectrum, we average the random sky noise spectra of those exposures and cameras, denoted $\left< \delta F_{sky} \right>_q$ in what follows, and then we replace the original \lya\ $\delta_q$ by this average sky noise spectrum divided by the quasar continuum used to derive the $\delta_q$ and including the mean flux transmission fraction. In other words, we have

\begin{equation}
\delta^{s}_q(\lambda) \equiv \frac{ \left< \delta F_{sky}\right>_q }{C_q \overline{F}} (\lambda).\label{eq:delta_sky}
\end{equation}

The next step is to apply the projection\footnote{The projection is a linear operation that is distributive and applies to each additive component of $\delta_{q}$, among which the sky noise term $\delta_{q}^{s}$.} that removes the mean and slope from each line of sight  to get the $\hat{\delta}^{s}_q$. This step is part of the estimator of the correlation function \citep{dMdB2021,RamirezPerez2023}. In our case, it emulates the effect of continuum fitting. \\

\begin{figure}[!]
\centering
\includegraphics[width=0.75\columnwidth]{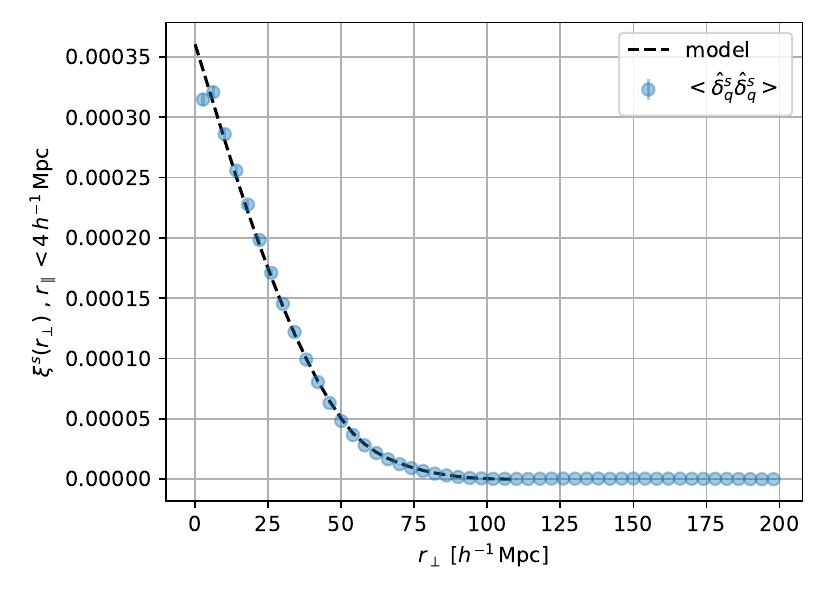}
\includegraphics[width=0.75\columnwidth]{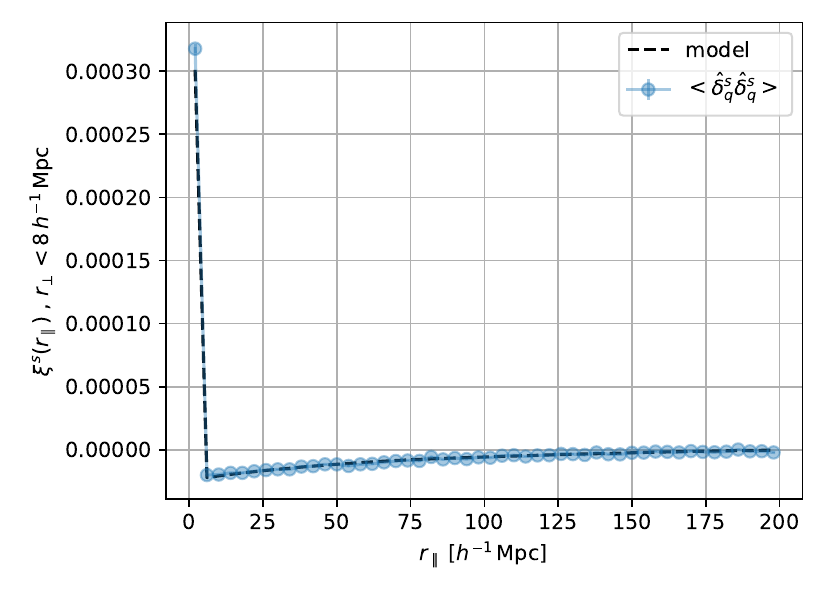}
\caption{Expected contribution of the sky model noise to the DESI Y1 \lya\ auto-correlation. The blue dots are the measured correlation function from the mock $\delta^{s}_q$ data set which emulates the sky model noise contribution to the \lya\ forests, and the dashed curve is the model described in \S\ref{sec:sky-model} derived from the geometry of a DESI petal and accounting for the distortion induced by continuum fitting. The top plot is the correlation as a function of the transverse separation $r_\perp$ for the first $r_\parallel$ bin which includes pairs at the same wavelength and hence directly the white noise contribution. The bottom plot is the correlation as a function of $r_\parallel$ averaged over the first two bins of $r_\perp$ highlighting the effect of continuum fitting for values of $r_\parallel>0$.}
\label{fig:skynoise}
\end{figure}

The auto-correlation of this mock data set $\{\hat{\delta}^{s}_q\}$ gives us the contribution from the sky noise to the \lya\ correlation function. It is represented on Figure~\ref{fig:skynoise}.
One can see that the signal is mostly limited to the first $r_{\parallel}$ bin as expected from this white noise simulation (i.e. simulation of the uncorrelated noise between pixels). The extension to larger values of $r_{\parallel}$ is entirely caused by the distortion due to the effect of continuum fitting (actually the projection for our simplified mock data set).\\

Along $r_{\perp}$, one can see that the signal does not extend beyond $\sim 100$ \mpc. It is as expected. We derive a more predictive model in the next section.\\

It is worth noting from Eq.~\ref{eq:delta_sky} that the amplitude of this sky noise correlation function will decrease with increasing survey depth (or number of observations per target for DESI) and the target brightness. The data shown here is expected to be a fair estimate of the amplitude of the signal in the DESI Y1 \lya\ auto-correlation.

\subsubsection{Modeling the noise correlation function}
\label{sec:sky-model}

Because the sky model noise is not correlated as a function of wavelength in first approximation -- at least it is the case in the emulation presented above -- any non-null correlation for $r_\parallel > 0$ must be the consequence of the continuum fitting. As mentioned in Section \ref{sec:lya}, this effect on the correlation function is characterized by a distortion matrix $D$ which applies to the model of the undistorted 2D correlation function when represented as an array of values in bins of $(r_\parallel,r_\perp)$. The undistorted noise correlation function is of the form $\xi_{noise} (r_\parallel,r_\perp) = \delta^K(r_\parallel) f(r_\perp)$. The function $f(r_\perp)$ is proportional to the probability that at a pair of measurements separated by $r_\perp$ is coming from spectra observed in the same petal and exposure. It can be derived from the geometry of a petal and the fiducial cosmology used to convert angles and redshifts into distances.\\

A DESI petal is represented in Figure \ref{fig:desi-petal} with the circular patrol area of its 500 fiber positioners. The probability of a pair of random coordinates to be accessible by fibers in the same petal can be computed numerically for this data, picking an arbitrary normalization. The angular separation can be converted to comoving separation provided a fiducial cosmology (given above) and a fiducial redshift, here $z_{fid}=2.4$. The difference between this value of $z_{fid}$ and the effective redshift of the final DESI Y1 sample is smaller than 0.1. This is close enough for the level of precision required with the current data sample.\\

We choose to normalize the model such that $\xi_{noise}(r_\parallel=0,r_\perp=0) \equiv 1$ and fit a multiplicative amplitude $a_{noise}$ to the measurements.
This gives the model represented as a black dashed curve in the top panel of Figure~\ref{fig:skynoise}, with a fitted amplitude $a_{noise} \simeq 3.8\times 10^{-4}$. The bottom panel shows the additional effect of the distortion matrix. The figure highlights the excellent agreement between the model and the mock data. In an earlier work,~\cite{Gordon2023} used a second order polynomial of $r_\perp$ instead of the numerical computation presented here. Both forms are close but we recommend using the more accurate numerical result of this paper in future works. It is included in the \texttt{vega} software\footnote{\url{https://github.com/andreicuceu/vega}} used for the fit of DESI data.

\subsection{Spectrophotometric calibration uncertainties}\label{sec:cal}
\subsubsection{Description}

Spectro-photometric (or flux) calibration uncertainties introduce correlations in the quasar
spectra and therefore in the Lyman-$\alpha$ auto-correlation.
In DESI, as in previous surveys like BOSS/eBOSS, the flux calibration is derived from a comparison of the measured spectra of standard stars with stellar models (see \cite{Guy2023} \S4.8 and \S4.9). For DESI, this is performed independently from petal to petal as is the case for the sky subtraction. The precision of the fit of stellar models can be evaluated by comparing the measured colors (as a difference of magnitudes) of stars from the imaging survey used for targeting with the synthetic colors derived from the stellar models. We find in DESI a typical RMS value of 0.02 in $g-r$ per star\footnote{$g$ and $r$ are SDSS-like filters used in the imaging surveys from which the DESI targets were selected, see~\cite{Dey2019}.}. Given that about 10 fibers are pointed to standard stars per petal, this corresponds, after averaging, to a calibration uncertainty of about 0.6\% at a wavelength scale of about 2000 \AA, which is the typical width of a filter. Because we fit a normalization and slope per forest, as explained in Section \ref{sec:lya}, any calibration error at a scale larger than 500 \AA\ is suppressed. On the other hand the contribution of the measurement white noise to the calibration, at a scale given by the spectral pixel size of 0.8 \AA, would appear like the sky noise described in the previous section. Thus we are concerned by the noise introduced at intermediate scales and caused by modeling errors of the standard stars absorption lines, or on the contrary, caused by variations of the throughput during the night and from fiber to fiber. Of concern, for instance, is the variation with humidity of the position of a transmission dip in the spectrographs collimator mirror reflectivity around 4400 \AA\ (\cite{desi-instrument-overview-2022}).

\subsubsection{Characterization}

We aim at estimating the flux calibration errors while preserving their correlation across wavelengths. \\

Calibration vectors  $C_{p,e}(\lambda)$ (as defined by Eq.~17 of~\cite{Guy2023}) translate flat-fielded and sky-subtracted fiber spectra in units of electrons$\cdot$\AA$^{-1}$ into spectral energy densities in units of ergs$\cdot$s$^{-1}$$\cdot$cm$^{-2}$$\cdot$\AA$^{-1}$.
They are determined for each petal $p$ and exposure $e$,  from the ratio of the measured counts to expected flux of calibration stars. We use the average ratio over all of the calibration stars measured in the same petal. About 10 calibration stars are observed per petal. The exact calibration procedure is more complex as it accounts for variations of spectral resolution from one fiber to the next and corrections for variations of fiber acceptance in the focal plane which vary very slowly with wavelength and can be ignored here (see~\cite{Guy2023} for more details). \\

Our approach is to use the relative variation of the flux calibration vector $C_{p,e}(\lambda)$ as a measurement of calibration errors.
Our objective is to isolate a signal that we think to be null if the calibration step has been done optimally, and non-zero if there are any calibration errors. We will compare different petals and exposures from different tiles and correct for the average effects arising from atmospheric corrections and temporal effects arising from humidity variations, as well as variations from spectrograph to spectrograph.\\

For that, we start by defining the contrast between the flux calibration vector of a petal and the average over all 10 petals. When comparing different exposures, we only consider exposures of different tiles so as to get a different set of calibration stars. It follows that the variation of calibration from one exposure to the other will account for the errors arising from the stellar model used.

\begin{equation}
   \delta C_{p,e}(\lambda) = C_{p,e} / \left< C_{p,e} \right>_{p} (\lambda) -1
   \label{eq:cal}
\end{equation}
This quantity removes, to a good approximation, any genuine change in the calibration with exposure time, airmass, sky transparency and seeing.\\

Then, we look at how $\delta C_{p,e}$ varies from one exposure to the next. The purpose of our approach is to reduce the contribution of the known time-dependent variation of calibration with humidity. As a result, that contribution is minimized when looking at successive exposures. Therefore, the quantity of interest for our study is:

\begin{equation}
  \delta C^{\prime}_{p,e}(\lambda) = \frac{1}{\sqrt{2}} \left( \delta C_{p,e} - \delta C_{p,e+1} \right)(\lambda) .\label{eq:cal2}
\end{equation}

\begin{figure}[!]
\centering
\includegraphics[width=0.75\linewidth,angle=0]{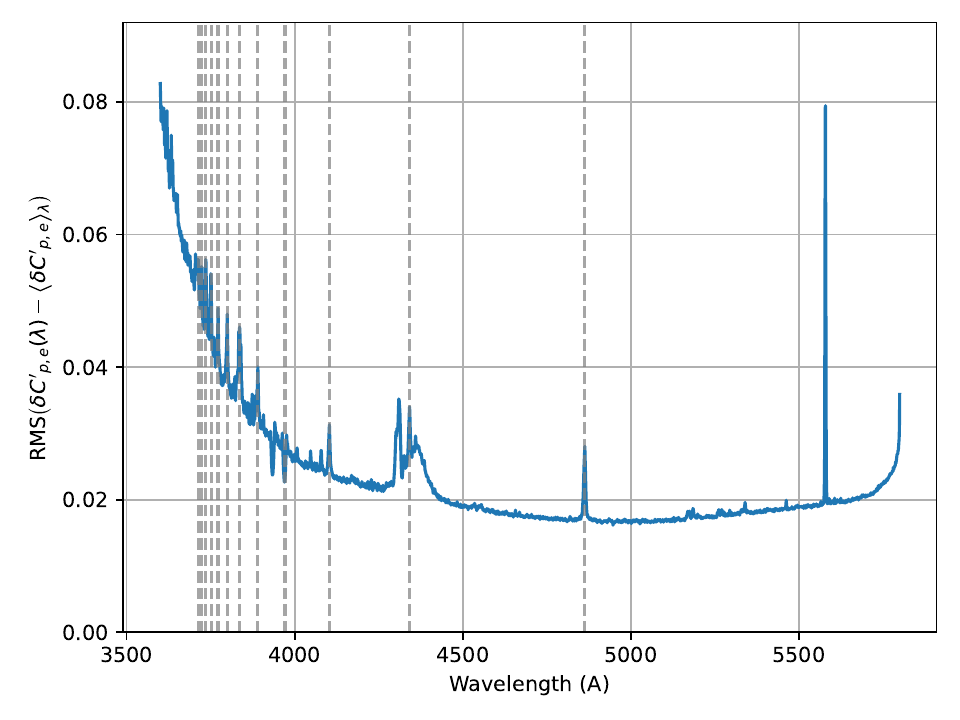}
\caption{RMS of the spectrophotometric calibration residuals $\delta C^{\prime}_{p,e}(\lambda)$ for the blue cameras after subtracting an average over wavelength for each exposure ($e$) and camera ($p$ for the petal associated to the camera). The latter is done in order to mimic, to first approximation, the effect of continuum fitting.
The vertical dashed lines indicate the wavelengths of the Balmer series.}
\label{fig:dcal-rms}
\end{figure}

Let us study the covariance of that quantity. The RMS of the $\delta C^{\prime}$ are shown as a function of wavelength in Figure \ref{fig:dcal-rms} after subtracting an average over wavelength for each exposure and camera as a first approximation to account for the continuum subtraction. The calibration error is larger at shorter wavelength because of the lower throughput, which results in a larger measurement noise. On top of that, several spikes are present. The largest spike at 5579 \AA\ is caused by the oxygen sky line which is masked out. The broader excess around 4350 \AA\ is caused by the transmission dip. On one hand, the transmission dip naturally induces a higher relative measurement noise caused by the lower throughput at this wavelength. On the other hand, there are actual variations of the throughput from one exposure to the next. One can also see several sharp features that correspond to the Balmer line series and are marked as vertical dashed lines in Figure \ref{fig:dcal-rms}. Those excesses are unambiguously calibration errors. They can be caused by measurement uncertainties in the star spectra simply because the signal to noise is lower at those wavelengths. It can also be caused by errors in the modeling of the stellar atmospheres, a mismatch between the observed and expected resolution of the instrument, or some noise in the measurement of the radial velocities of stars.\\

We now compute the 1D correlation function ($\xi^c_{1D}$) of the calibration errors as a function of the comoving separation along the line of sight $r_\parallel$. We can convert wavelength differences into comoving separations using the \lya\ wavelength for redshifts and our fiducial cosmology. This 1D correlation function is given by
\begin{equation}
\xi^c_{1D}(r_\parallel) =  \left< \, \delta C^{\prime}_{p,e}(\lambda) \, \delta C^{\prime}_{p,e}(\lambda + \Delta_\lambda( \lambda, r_\parallel)) \, \right>_{p,e,\lambda} \label{eq:xi1d}
\end{equation}
where the average is over all petals $p$, exposures $e$, and wavelength $\lambda$, and where $\Delta_\lambda(\lambda, r_\parallel)$ is the wavelength difference that corresponds to the comoving separation $r_\parallel$ at the wavelength $\lambda$.\\

\begin{figure}[!]
\centering
\includegraphics[width=0.75\linewidth,angle=0]{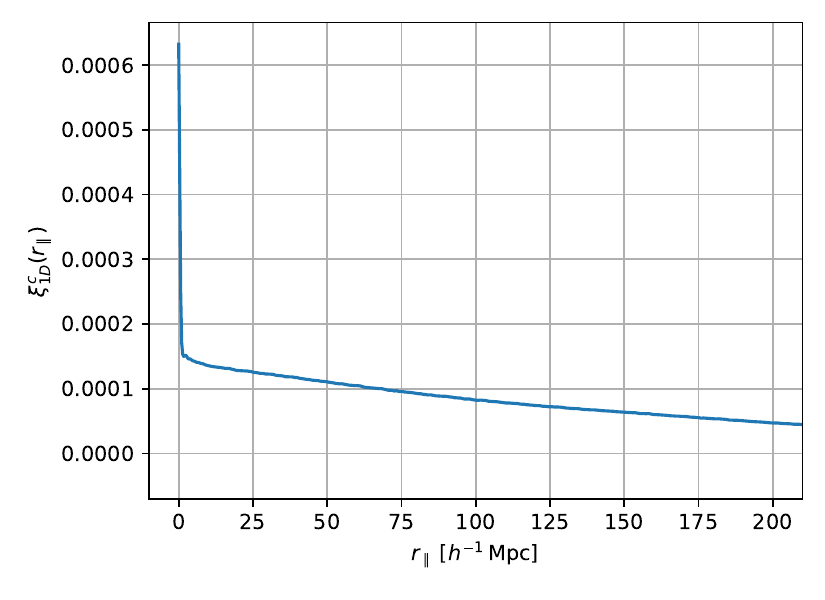}
\caption{Average one dimensional auto-correlation function $\xi^c_{1D}(r_\parallel)$ of the spectro-photometric calibration errors $\delta C^{\prime}_{p,e}(\lambda)$ (see equation~\ref{eq:xi1d}).}
\label{fig:dcal-cf-1d}
\end{figure}

The result is shown in Figure~\ref{fig:dcal-cf-1d}. Apart for the peak at 0 separation which corresponds to a white noise term with an RMS of about 0.02, the correlation function is featureless. The shape as a function of $r$ will be different once we account for the  \lya\ forests specific length, weights and the effect of the projection. Its amplitude will also be reduced when averaged over several exposures.\\

\subsubsection{Calibration noise 2D correlation function}

Similarly to the sky subtraction test presented in \S\ref{sec:sky-noise-characterization}, we use the actual DESI Y1 \lya\ forest data set, keeping all properties including weights but, replacing the $\delta_q$ values by random calibration errors. We use directly one of the $\delta C^{\prime}_{p,e}(\lambda)$ array for each petal $p$ and exposure $e$ instead of picking random Gaussian values in order to preserve the correlation across wavelengths. More precisely, for each \lya\ line of sight, we identify the list of exposures $\{e\}$ and petals $\{p\}$ that were used to produce the combined quasar spectrum in which the forest is measured, and then we replace the original \lya\ $\delta_q$ by the average of the calibration errors.

\begin{equation}
\delta^{c}_q(\lambda) \equiv \left< \delta C^{\prime}_{p,e}\right>_q (\lambda)
\end{equation}

\begin{figure}[!]
\centering
\includegraphics[width=0.75\linewidth,]{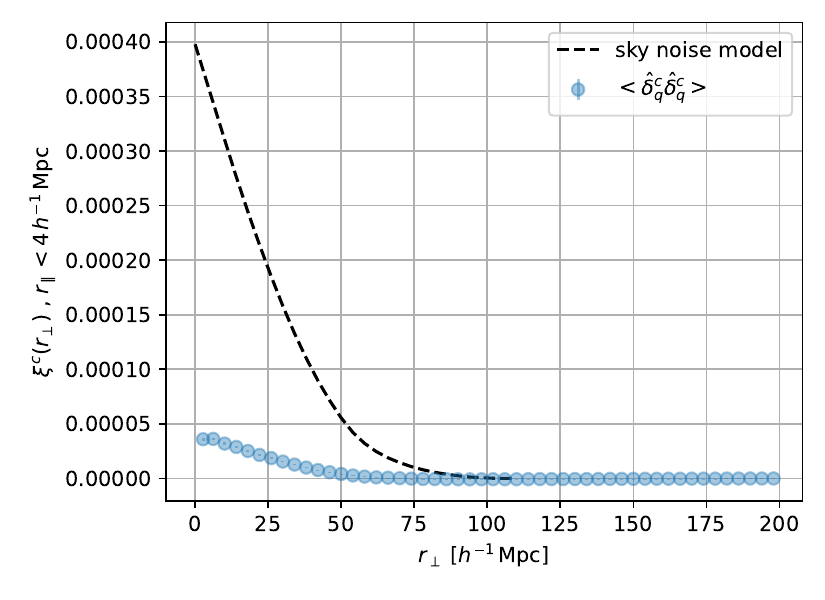}
\includegraphics[width=0.75\linewidth,]{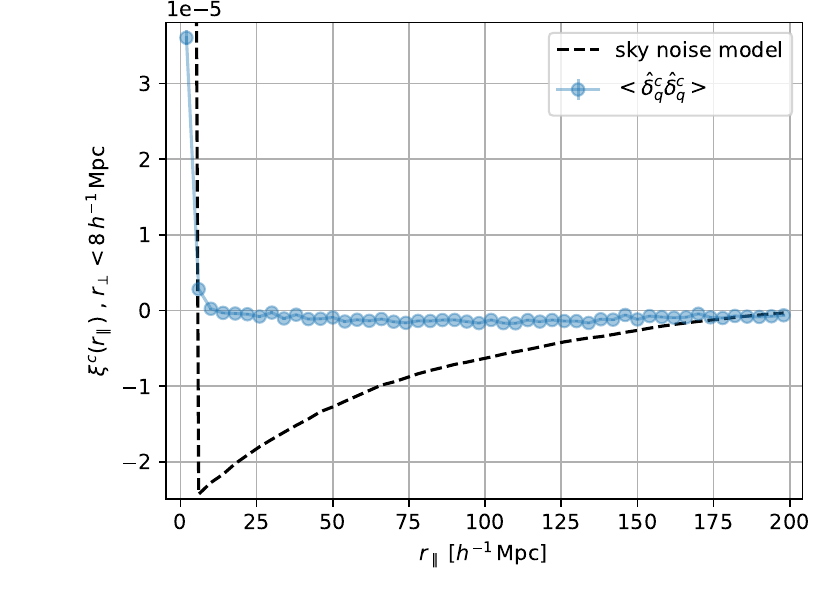}
\caption{Expected contribution of the spectro-photometric calibration noise to the DESI Y1 \lya\ auto-correlation. The blue dots are the measured correlation function from the mock calibration noise contribution $\delta^{c}_q$, and the dashed curve is the model described in \S\ref{sec:sky-model} for the sky model noise for comparison. The top plot is the correlation as a function of the transverse separation $r_\perp$ for the first $r_\parallel$ bin, the bottom plot is the correlation as a function of $r_\parallel$ averaged over the first two bins of $r_\perp$.
\label{fig:calibnoise-cf-2d}}
\end{figure}

In the same way that we did for the sky, we apply the projection to obtain the $\{\hat\delta^{c}_q\}$. Then, we measure their 2D auto-correlation function. The result is shown in Figure~\ref{fig:calibnoise-cf-2d}, where one can see that the contribution of the calibration noise is much smaller than that of the sky model noise. In the lower plot, one can note that the shape of the correlation as a function of $r_\parallel$ does not follow that of the sky noise, indicating some genuine correlation as a function of wavelength, but this contribution is very small and can safely be neglected for the \lya\ analysis. As an example, at $r_\parallel \sim 100$~\mpc, the amplitude of the \lya\ auto-correlation function is about $2 \times 10^{-4}$, when that of the calibration noise is of $2 \times 10^{-6}$. Subtracting this contamination to the \lya\ auto-correlation function (both in the Lyman-$\alpha$ and Lyman-$\beta$ forest regions) in the DESI Y1 cosmology fit \citep{KP6} changes the best fit BAO parameters by less than 0.05\% and increases the combined $\chi^2$ of the fit by about one\footnote{This is the increase in best fit $\chi^2$ of the fit that combines the \lya\ auto-correlation functions and the \lya-QSO cross-correlations with the \lya\ aborption measured both in the Lyman-$\alpha$ and Lyman-$\beta$ forests, called regions A and B in \cite{KP6}.}.\\

\subsection{Comparison with BOSS and eBOSS}

The SDSS focal plane instrument used for both the BOSS and eBOSS surveys is of comparable angular size to DESI, with a field of view angular diameter of about 3~deg~\citep{Smee2013}. The focal plane fibers are separated in two halves, each feeding a different spectrograph. As for DESI, the data processing is performed independently for each spectrograph, and this includes the sky and calibration models.\\

\cite{Bautista2017} and \cite{dMdB2020} analyzed the BOSS and eBOSS \lya\ correlation functions and also found that the noise correlations were dominated by the sky subtraction noise. \cite{Bautista2017} find a noise correlation in the CIV forest of $2 \times 10^{-4}$ in the first separation bin ($r_\perp,r_\parallel < 4\,$\mpc, see their Figure 8) and \cite{dMdB2020} find a value of about $5 \times 10^{-5}$ in the same bin in the MgII region (see their Figure 9). Those differences are explained by different relative brightness of the quasars compared to the sky level, and variations in the number of fibers allocated for the fit of the sky background. DESI is targeting fainter quasars and only a fraction of them have been observed at full depth in the first year sample, so it is natural that we find a larger contribution of the sky model noise in the DESI Y1 data set.

\subsection{Summary}

We have identified two sources of correlated noise coming from the data processing pipeline: the sky background model noise and the spectro-photometric calibration noise. The effect of both of those terms on the DESI Year 1 \lya\ auto-correlation function have been evaluated. A numeric model of the contribution of the white noise has been proposed. We also find that the component of the calibration noise that is not white noise can be safely neglected in the analysis of the \lya\ auto-correlation function. However, we have not modeled all of the possible sources of correlated noise, and specifically for the sky noise, we have not addressed all the sources of noise that are not coming from the model itself. In the following section, we will measure the correlation function of the flux transmission fraction of low redshift quasar spectra, where there is no \lya\ forest absorption in order to estimate the combined contribution of all of the sources of correlated noise plus the auto-correlation of foreground absorbers.

\section{Astrophysical contaminants}
\label{sec:astro}

In order to validate the DESI noise correlation function model derived in the previous section, and verify whether it is sub-dominant compared to the cosmological signal, one wants to make a direct and isolated measurement of the spurious correlations arising from the instrument.
That requires devising some sort of a null test, i.e. correlating parts of the quasars' spectra that contain a minimal amount of cosmological signal while retaining all of the instrumental noise.\\

The difficulty with this comes from the fact that hydrogen is not the only absorber with which the light from the quasar interacts while crossing the universe on its way towards us.
Indeed there are irreducible astronomical foreground in the form of metal absorption that make any attempt at a real null test unattainable.
As a result, we resort to study the auto-correlations in metal spectral regions of the quasar rest-frame spectrum, redward of the \lya\ forest.
This gives us the opportunity to measure the combined contamination from correlated noise and foreground absorbers.

\subsection{Metal spectral regions}\label{sec:metals}

\begin{figure}
\centering
\includegraphics[width=0.98\linewidth,angle=0]{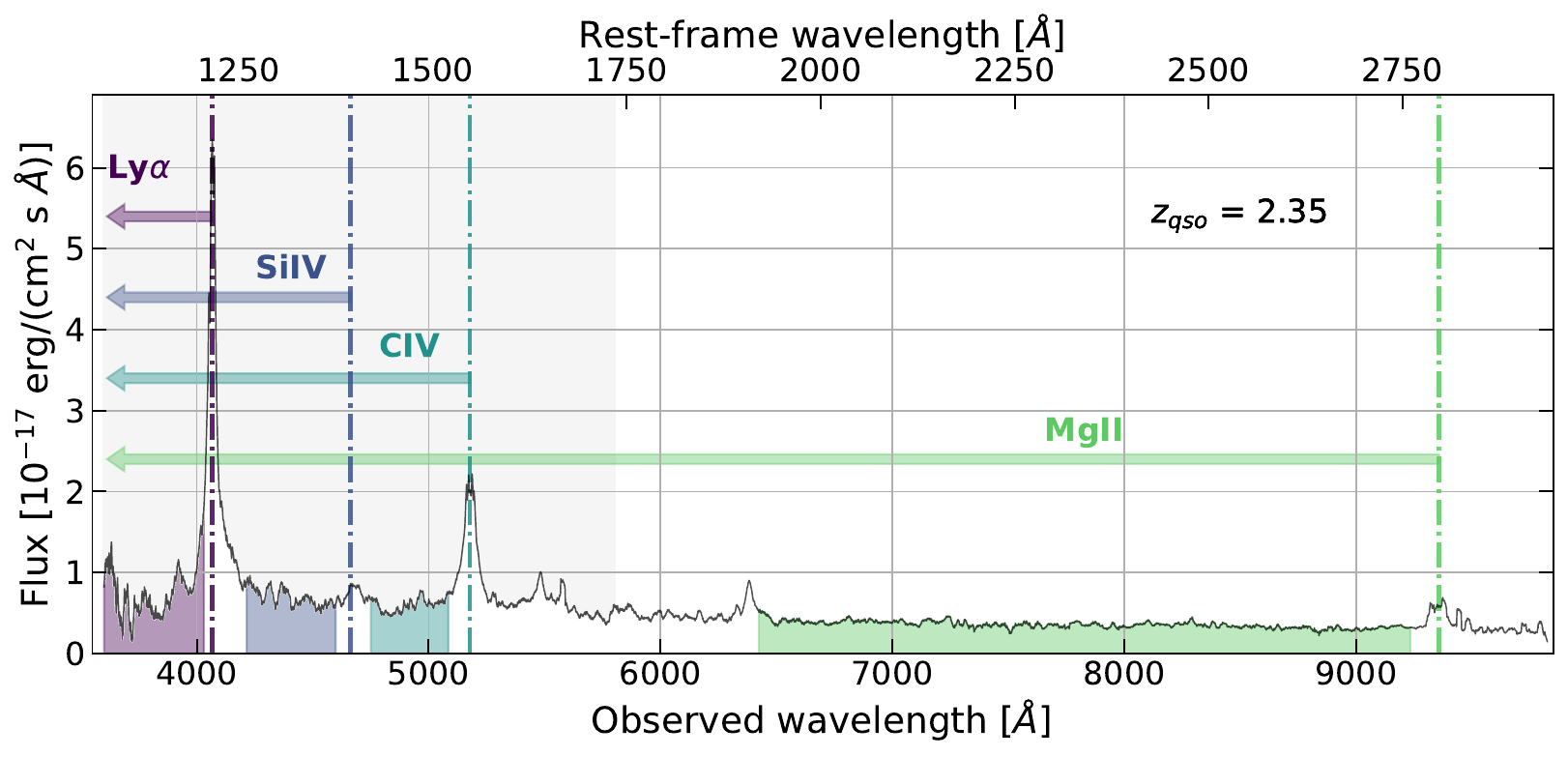}
\caption{Quasar spectrum at $z=2.35$ displaying several emission lines and the \lya\ forest. The \lya\ forest is highlighted in purple. The MgII, CIV and SiIV forests are highlighted in various shades of green to blue. MgII absorption span leftward of the MgII peak, contaminating CIV, SiIV, and \lya\, forests. CIV absorption span leftward of the CIV doublet, contaminating SIV and \lya\, forests. SiIV absorption span leftward of the SiIV peak, contaminating the \lya\, forest. The blue spectrograph region is shaded in gray. }
\label{fig:sidebands}
\end{figure}

In the rest-frame of a high redshift quasar, foreground absorption features from a given chemical species will appear at shorter wavelength than the corresponding rest-frame emission peak.
The metal with the highest rest-frame wavelength will therefore contaminate the regions affected by other metal absorption at smaller rest-frame wavelength, as highlighted on Figure \ref{fig:sidebands}.
We start by characterizing the auto-correlation of the MgII forest, then the CIV forest and finally the SiIV forest, going from the most pristine region towards increasingly contaminated regions.

\begin{table}[h!]
  \small
  \centering
  \begin{tabular}{ccc}
Spectral region & $\lambda_{RF}$ range & QSO redshifts\\
\hline
MgII & 1920-2760\,\AA & 0.3-2.0\\
CIV  & 1420-1520\,\AA & 1.4-3.1\\
SiIV & 1260-1375\,\AA & 1.6-3.6\\
\lya\ & 1040-1205\,\AA & 2.0-4.5\\
\end{tabular}
\caption{
The spectral regions we define and their corresponding quasar rest-frame wavelength ranges.
 The quasar redshift range is derived for the fixed common observer-frame wavelength range 3600-5772\AA\ (we omit here the additional requirements on the minimal length of a forest which restrict further the redshift range).\label{table:metals-prop}}
\end{table}

We summarize the properties of each metal spectral region in Table~\ref{table:metals-prop}. The boundaries of the MgII forest are set by the CIII emission line at 1908\AA\ and the MgII emission line at 2796\AA\, which means we will not get contamination from any IGM absorption at a wavelength shorter than 1920\AA\ in this sample but we expect contamination from larger wavelengths, including MgII. The CIV forest is bounded by the Si~IV 1375\AA\ emission line and the CIV 1548\AA\ emission line. It receives contributions from any absorber at $\lambda_{RF}>1375$\AA, we expect its auto-correlation to be dominated by CIV absorption, with minor contributions from AlII, AlIII, FeII, MgII (see~\cite{Pierietal2014,Yang2022,Morrison2023}). Finally, the Si~IV forest is bounded by the onset of intervening SiII 1260\AA\ absorption and the SiIV 1394\AA\ emission line. It receives contributions from the same species as the CIV forest, plus OI, SiII, CII and the SiIV lines that we expect to be sub-dominant only to CIV absorption.
We note that the SiII 1260\AA\ and other Si lines at shorter wavelength are detected unambiguously and included in the fit of the \lya\ auto-correlation as they appears clearly in their cross-correlation with the \lya\ absorption (see e.g. ~\cite{dMdB2020}).

\subsection{Correlation functions}\label{sec:metals2}

For each forest, we convert wavelength to redshifts considering the \lya\ line in order to emulate correctly how the absorbers contaminate the \lya\ forests. This means that we will need to use the metal matrices introduced in \S\ref{sec:lya-metal-mat} that encode the transformation from the correct co-moving separation of a pair of absorbers to the incorrect co-moving separation one derives when using the \lya\ line to convert wavelength to redshifts.\\

\begin{figure}[!]
\centering
\includegraphics[width=0.75\linewidth,angle=0]{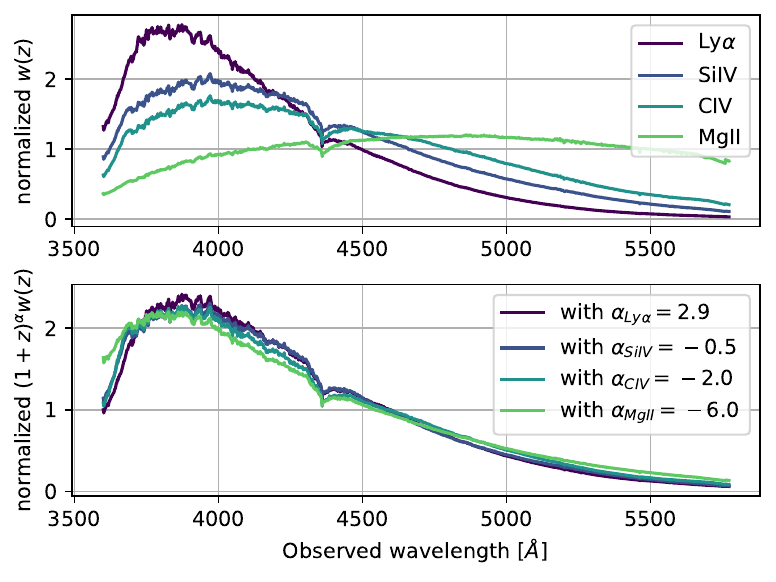}
\caption{Top: normalized distributions of weights $w(z)$ as a function of wavelength (or equivalently redshift) for the various forests MgII, CIV, SiV and \lya. Bottom: same figure with scaled weights $w^{\prime}(z) \propto (1+z)^\alpha \, w(z)$, with $\alpha$ values chosen such that the $w^{\prime}$ weighted average redshift is $\left< z_{Ly\alpha} \right> = 2.35$ for all the wavelength regions.}
\label{fig:weights-vs-z}
\end{figure}

We also make sure to use the same distribution of weights as a function of wavelength as in the \lya\ forest in each metal spectral region, in order to avoid any bias in the integrated correlation function caused by a redshift evolution of the clustering strength of the absorbers. We use the same terms entering in the definition of the weights as for the DESI Y1 \lya\ forest (same values of $\eta$ and $\sigma_{LSS}$ in Eq.~6 of \cite{RamirezPerez2023}), but we apply different scaling as a function of redshifts in order to compensate for the changes of signal to noise with wavelength for the different quasar samples.  The average weights as a function of redshift, before and after this scaling, for all the forests are shown in Figure~\ref{fig:weights-vs-z}.\\

\begin{figure}[!]
\centering
\includegraphics[width=0.75\columnwidth]{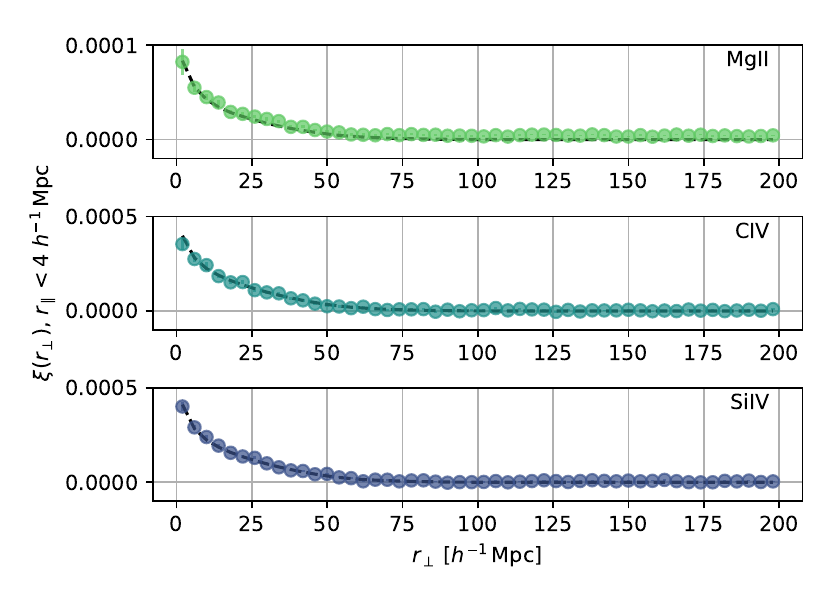}
\includegraphics[width=0.75\columnwidth]{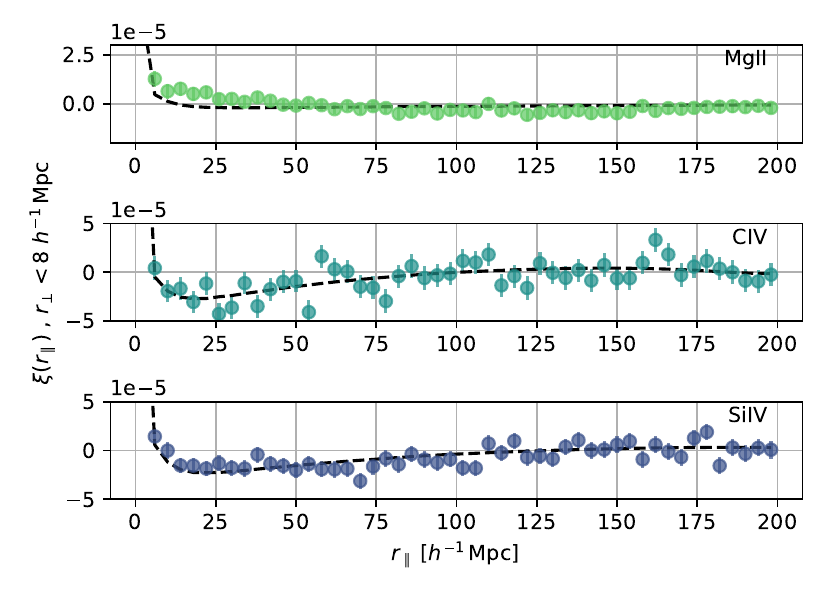}
\caption{Correlation functions of the MgII, CIV and SiIV forests (colored dots with error-bars) along with their best fit model (black dashed curve). The top plot is the correlation as a function of the transverse separation $r_\perp$ for the first $r_\parallel$ bin, the bottom plot is the correlation as a function of $r_\parallel$ averaged over the first two bins of $r_\perp$ (note that the values in the first $r_\parallel$ bin are outside of the graph range in the lower plot). The best fit parameters are given in Table~\ref{table:metals-autocorr-bestfit}.
\label{fig:metals-cf-2d}}
\end{figure}

The auto-correlation function of the three forests are shown in Figure~\ref{fig:metals-cf-2d}. The SiIV and CIV correlations as a function of the transverse separation $r_\perp$ for the first $r_\parallel$ bin show a strong signal at small separation, the signal in the SiIV forest being the largest. The MgII correlation is weaker than the two others.

\subsection{Metal spectral regions auto-correlation function model and fit}
\label{sec:side-band-auto-correlations}

The metal spectral regions correlation functions are fitted with a two component model. The first one is the noise correlation function model $a_{noise} \,\xi_{noise}$ described in \S\ref{sec:sky-model}, and the second one is the combined signal from the auto-correlation functions of the foreground metal absorbers (hereafter $\xi_{m}$). We consider either a single (effective) absorber or try to isolate the contribution from several of them.

$\xi_{m}$, for a metal absorber '$m$', is obtained from the Fourier transform of a power spectrum model ($P_{m}(k,\mu_k)$, described below), and is then remapped from the true comoving separation (based on the redshifts $z=\lambda/\lambda_{m}-1$, where $\lambda_{m}$ is the metal absorber wavelength) to the assumed comoving separations (based on the redshifts $z=\lambda/\lambda_{Ly\alpha}-1$) with a precomputed metal matrix $M_{m}$ (see~\S\ref{sec:lya-metal-mat}).
Both components are then once again transformed with the distortion matrix $D$ to account for the continuum fitting.
Schematically, the correlated signal in a metal spectral region is modeled as
\begin{equation}
  \hat{\xi} = D [ a_{noise} \, \xi_{noise} + \sum_{m} M_{m}\cdot\xi_{m} ]   \label{eq:xi-sb-sky-metal}
\end{equation}

The power spectrum model for the metal absorber $m$ is

\begin{equation}
  P_{m}(k,\mu_k) = b_{m}^2 ( 1+\beta_{m} \mu_k^2)^2 P_{QL}(k) \label{eq:Pciv}
\end{equation}

Following the approach of~\cite{dMdB2020}, the quasi linear matter power spectrum $P_{QL}(k)$ is derived from the linear matter power spectrum $P_{lin}(k)$ by decoupling and broadening the BAO peak (see their Eq.~28). $P_{lin}(k)$ is precomputed with CAMB \citep{CAMB}, for the fiducial cosmology at a fixed reference redshift $z_{ref}$. $P_{QL}(k)$ is then multiplied by a Kaiser term for the redshift space distortions, with the bias parameters $b_{m}$ and $\beta_{m}$. Note that formally, we do not need to include the broadening of the BAO peak in our study since we are focusing on smaller scales. However, in the interest of consistency, we made the choice to use the same model for metals as used in the main \lya\, analysis \citep{KP6}.\\

We now provide details for the metal absorbers considered in each of the metal spectral regions.
\begin{itemize}
\item For the MgII spectral region, we consider only the MgII absorption and adopt a single effective rest-frame wavelength $\lambda_{MgII} = 2799$\AA\ to account for the combined effect of the absorption occurring at 2796\AA, 2804\AA, and 2853\AA\ (see \cite{dMdB2020}). As we will see in the next section, the auto-correlation of the MgII spectral region is probably contaminated by other foreground absorbers, so we will fit for an effective bias $b^{eff}_{MgII}$, with the subscript MgII to note that we use the MgII wavelength to compute the metal matrix, but keeping in mind that this bias is probably the result of the combined effect of multiple absorbers.

\item For the CIV spectral region, we consider an effective CIV rest-frame wavelength $\lambda_{CIV} = 1549.1$\AA. It is actually a doublet of lines at $1548.2$\AA\ and $1550.8$\AA. As previously done in \cite{Gontcho2018} and \cite{Blomqvist2018} we do not try to differentiate them here. We consider only this absorber in the fit because it dominates the others, but the CIV spectral region auto-correlation function also receives contributions from MgII and other foreground absorbers. As a consequence, we will fit for a single effective bias $b^{eff}_{CIV}$.

\item Finally, for the SiIV spectral region, we also consider for our baseline the CIV absorption, fitting again an effective bias $b^{eff}_{CIV}$ that will this time receive an extra contribution from the SiIV absorption. We also test the possibility to model separately the auto-correlation of the SiIV, considering for this an effective wavelength of 1396.8\AA\ (when it is actually a doublet of lines).
\end{itemize}

For all metal absorbers, we use a redshift space distortion parameter $\beta_m=0.5$ following \cite{Bautista2017}.

\subsection{Fit results}

The fit parameters are the amplitude $a_{noise}$ of the noise correlation function and the bias parameters for the absorbers.
The best fit parameters and reduced $\chi^2$ values are given for all three metal spectral regions in Table~\ref{table:metals-autocorr-bestfit}. We note that the measured biases and correlated noise amplitudes  are quite correlated, with a correlation coefficient close to 0.5. We also obtain different values of the correlated noise amplitude $a_{noise}$ between the spectral regions. This is caused by differences in the continuum level that modulates the sky residual noise (see equation~\ref{eq:delta_sky} where the sky residuals are divided by the continuum).\\

\begin{table}[h!]
\renewcommand{\arraystretch}{1.3}
  \centering
  \begin{tabular}{ccccc}
Spectral region &  bias &  $a_{noise} \times 10^4$ & $\chi^2 / ndf $ & $ndf$ \\
\hline
Mg II &  $b^{eff}_{MgII} = -0.0053 \pm 0.0005$              & $0.42 \pm 0.01$  & 1.54 & 1588 \\ 
CIV  &  $b^{eff}_{CIV} = -0.0182 \pm 0.0015$ & $2.73 \pm 0.09$ & 1.06 & 1588 \\ 
SiIV &  $b^{eff}_{CIV} = -0.0191 \pm 0.0012$ & $2.57 \pm 0.04$ & 1.11 & 1588 \\ 
\end{tabular}
\caption{Auto-correlation function fit parameters for the metal spectral regions. $\delta_q$ weights have been rescaled as a function of wavelength in order to obtain for all forests the same weighted mean wavelength as for the \lya\ forest.\label{table:metals-autocorr-bestfit}}
\end{table}

We measure a relatively small but significant signal in the MgII spectral region. The fit result gives an effective bias that is larger than the MgII bias derived from the cross-correlation with quasars. \cite{dMdB2020}  obtain from the cross-correlation a combined MgII bias\footnote{This is the sum of the biases of the three MgII transitions reported in table 3 of \cite{dMdB2019} for the cross-correlation with quasars.} of $-0.0025 \pm 0.0002$ while we measure from the auto-correlation a bias $b^{eff}_{MgII} = -0.0053 \pm 0.0005$.
We also note that the reduced $\chi^2$ of our fit is larger than one. This suggests that the model is incomplete and that there are other contributions to the MgII spectral region signal. We do not try to identify them  in this paper because it is a very small signal (four times smaller that the combined effect of CIV and SiIV), but we note that this measurement provides us with a good estimate of the contribution of unidentified contaminants to the \lya\ forest (among which are interstellar medium absorbers for instance, see \cite{Vadai2017}).\\

For the CIV and SiIV spectral regions, we obtain better reduced $\chi^2$ because the CIV and SiIV signals dominate and the uncertainties are larger. The latter is due in part to the contribution of CIV and SiIV to the signal variance, and in part to the length of the forests and quasar redshift distributions.
 In the CIV spectral region, we obtain a effective bias $b^{eff}_{CIV} = -0.0182 \pm 0.0015$.
One can estimate the true CIV bias by subtracting the contribution of MgII and other contaminants obtained with the fit of the MgII spectral region. To a good approximation, this consists in subtracting quadratically the two effective biases\footnote{The biases are added quadratically and not linearly because the cross-correlation of MgII and CIV does not contribute to the measured auto-correlation at small separation.} while taking into account the change of growth of structure between the redshifts of the two absorbers,
\begin{equation}
\left( b_{CIV} \right)^2 \simeq \left( b_{CIV}^{eff} \right)^2 -  \left( b_{MgII}^{eff} \right)^2  \frac{D^2(z_{MgII})}{D^2(z_{CIV})}
\end{equation}

where $D(z)$ is the large scale structure growth factor at the redshift $z$.
This gives $b_{CIV} = -0.016 \pm 0.002$, for $z_{MgII} = 0.455$ and $z_{CIV} = 1.63$.
This result is marginally compatible with the CIV bias measured from the cross-correlation of CIV with quasars. For instance, \cite{Blomqvist2018} report a value of $b_{CIV} (1 + \beta_{CIV}) = -0.0183 \pm 0.0014$ at $z = 2$ from the SDSS BOSS/eBOSS data sample. This gives $b_{CIV} = -0.012 \pm 0.001$ for our choice of $\beta_{CIV}=0.5$.\\

This difference of bias is only barely
significant, but if true, we are then left with two hypotheses: i) there are other sources of correlated noise in the DESI data set that are not accounted for with our simple model, ii) the cross-correlation of quasars and CIV absorption is driven not only by large-scale structure but also by an astrophysical connection between the quasar positions and  the CIV absorption strength.\\

\begin{figure}[h!]
\centering
\includegraphics[width=0.75\columnwidth]{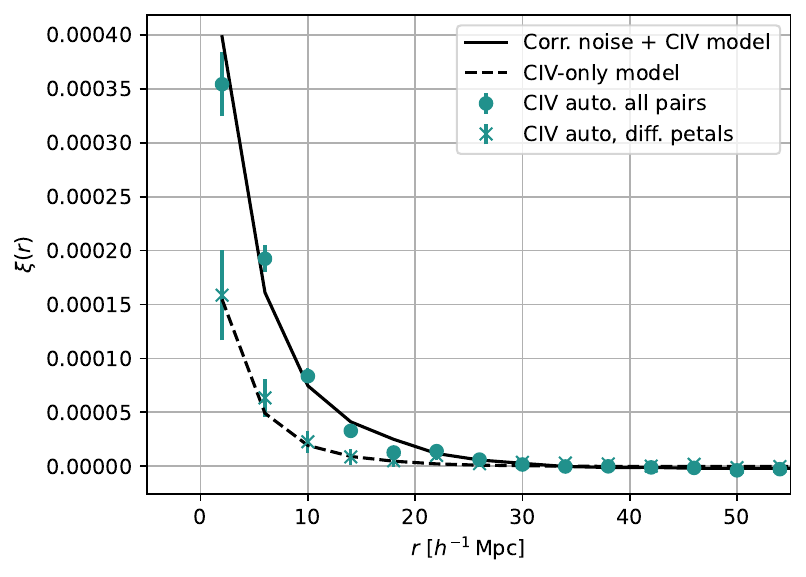}
\caption{Auto-correlation function as a function of separation in the CIV forest including all $\delta_q$ pairs (dots with error bars) or only those from different petals and exposures (stars), along with the best fit model from the fit of the full data set (solid curve, see Eq.~\ref{eq:xi-sb-sky-metal}), and the contribution from the CIV absorber correlation only (dashed curve, corresponding to the term $D M_m \xi_m$ in the same equation).
\label{fig:metals-cf-2d-different-petals}}
\end{figure}

In order to test the first hypothesis and verify that we have been able to separate the contribution of the noise correlations from that of genuine astrophysical sources, we repeat the measurement of the CIV forest auto-correlation by considering only $\delta_q$ pairs from quasar spectra recorded in different spectrographs (corresponding to different petals of the focal plane). This excludes most sources of noise coming from the data processing pipeline because most of the data reduction is performed independently for each spectrograph. The resulting correlation is compared to the measurement with the full data set in Figure~\ref{fig:metals-cf-2d-different-petals}. As expected, we find a much smaller signal when correlating spectra from different spectrographs and we further find that the measured correlation is consistent with the modeled CIV absorber auto-correlation (without the noise correlation). When we fit this correlation, we obtain $b^{eff}_{CIV} = -0.0183 \pm 0.0015$ and $a_{noise} = 0.4 \pm 0.1$. The bias is the same as with the full data set while the noise correlation is seven times smaller.
This result demonstrates that we are able to separate the instrumental noise correlation from the astrophysical signal in the metal spectral regions correlation fit, and that our noise correlation model is a good match to the data. As a  cross-check, we also measure the CIV auto-correlation function when considering only $\delta_q$ pairs from quasar spectra recorded in different nights, and we find a similar result. This additional test shows that there is no evidence of correlated noise coming from the daily calibrations  (CCD bias levels, spectrograph PSF, fiber flat fielding on the dome screen, see~\cite{Guy2023}) or any unanticipated variation of the instrument response from night to night.\\

We can also compare the amplitude of the noise correlation obtained in this fit of the CIV forest with the prediction presented in \S\ref{sec:sky} for the \lya\ forest. A fit of the emulated \lya\ forest sky noise correlation shown in Figure~\ref{fig:skynoise} gives a noise correlation amplitude $a_{noise}(\mathrm{Ly}\alpha) \sim 3.8 \times 10^{-4}$. Now we measure on average a larger quasar continuum in the CIV forest than in the \lya\ forest, with a ratio $C_{CIV}/C_{\mathrm{Ly}\alpha}$=1.13, so we expect
 a smaller noise correlation amplitude in the CIV forest after the flux is divided by the continuum. Numerically, this gives $a_{noise}(CIV) \sim a_{noise}(\mathrm{Ly}\alpha) ( C_{\mathrm{Ly}\alpha}/C_{CIV} )^2 \sim 3.0 \times 10^{-4}$. This is consistent within 10\% with the measurement reported in Table~\ref{table:metals-autocorr-bestfit}.\\

Concerning our second hypothesis, a natural source of correlation between the CIV strength and quasar positions is the inhomogeneity in the distribution of carbon ionizing photons arising from quasars. Quasars are thought to be a major source of the photons that maintain the IGM ionized but the mean free path of hydrogen ionizing photons is very large ($> 100\ h^{-1}$~Mpc, see \cite{Rudie2013}). On the other hand the mean free path of carbon (and silicon) ionizing photons can be very much shorter due to helium-II reionization effects \citep{Songaila1996,Giroux1997, Agafonova2007}. \cite{Morrison2021} probes the inhomogeneity of metal species such as CIV with respect to 3D proximity to quasars using eBOSS quasar spectra. They find some evidence of variation of the CIV bias with the distance from quasars though direct comparisons of their measurements and ours are non-trivial at this stage. We also note that \cite{Prochaska2014}, following the work of \cite{Tytler2009}, find a clear evidence of an excess of strong CIV absorbers in the vicinity of quasars to at least 1~Mpc. It is however difficult to extrapolate this result to the larger separations that we are studying here.
\\

Looking now at the SiIV spectral region, we measure a larger CIV effective bias $b_{CIV}^{eff} = -0.0191 \pm 0.0012$. A larger bias is expected because it includes the contribution of SiIV absorbers in addition to the CIV, MgII and other unidentified weak absorbers.
This increase in amplitude is consistent with the results on the 1D \lya\ power spectrum from the BOSS survey \cite{PalanqueDelabrouille2013} and the DESI early data release \cite{Ravoux2023}. Both references give a similar ratio of power between the SiIV region (their sideband SB2) and the CIV region (their sideband SB1). Considering the most recent result, one can see on Figure 9 from \cite{Ravoux2023} that this ratio is of $1.4 \pm 0.1$ at $k_{rest}=1\rm{\AA}^{-1}$ (with an uncertainty estimated from the variation of this ratio with $k_{rest}$ in the figure). This corresponds to a ratio of effective CIV bias of $1.2\pm 0.04$ (the square root of the previous number) which is consistent with our measured ratio of $1.05 \pm 0.1$ given the uncertainties.


We can not reliably isolate the SiIV contribution when looking at the combined auto-correlation without additional information. One possibility is to add a prior on the CIV effective bias $b^{eff}_{CIV}$ from the measurement obtained in the CIV forest. When refitting the SiIV forest with this method, we obtain $b_{SiIV} = -0.006 \pm 0.007$ when assuming a prior $b^{eff}_{CIV} = -0.0182 \pm 0.0015$. The quadratic sum of those two numbers, accounting for the change of growth factor, is consistent with the value obtained when fitting only the CIV effective bias. We also find that there is no gain in the quality of the fit (only 0.1 change in the fit $\chi^2$) when treating separately the SiIV absorber. In consequence, it is equivalent to consider one or two absorbers in the modeling of the auto-correlation function.
This measurement of a relatively weak SiIV bias is consistent with the analysis of \cite{Pieri2014letter} who found a CIV absorption about 5 times stronger than that from SiIV (and similarly MgII). They used a completely different technique based on stacked absorption profiles from absorbers detected in the CIV and SiIV regions of QSO spectra.

\section{Summary}
\label{sec:summary}

In this paper, we have studied the main contaminants of the \lya\ forest auto-correlation function measured with DESI. They can be separated in two broad categories: correlated residuals in the spectra introduced by the data processing pipeline, and astrophysical foreground absorbers in the intergalactic medium and interstellar medium (ISM). We have first characterized the correlated noise introduced by the data processing. The main contribution is the statistical noise from the sky background model that is subtracted to all spectra from the same exposure and spectrograph. We have estimated the signal level by emulating the noise in a mock data set, and we have proposed a simple model to account for it (see Fig.~\ref{fig:skynoise}). The second source of correlated noise comes from the spectrophotometric calibration uncertainties. We have characterized it by measuring the changes in calibration from one exposure to the next and one spectrograph compared to the others. We have found that this contribution is 10 times smaller and can be neglected (see Fig.~\ref{fig:calibnoise-cf-2d}). This measurement includes the calibration errors introduced by the fit of standard stars.\\

We have then measured the auto-correlation function of the transmitted flux fraction fluctuation ($\delta_q$) in various regions of the quasar rest-frame spectra (Fig~\ref{fig:sidebands}). We considered only forests with wavelength larger than the quasar \lya\ emission line; this ensures that the signal we measure is deprived from \lya\ absorption, but we have at the same time considered data in the same observer frame wavelength as the \lya\ forest, with the same relative weights as a function of wavelength.\\

We have looked at the auto-correlations of the MgII, CIV and SiIV spectral regions, fitting each of them with a two component model composed of the noise correlation model and a metal absorber auto-correlation function. We measure increasing values of the absorber bias for each spectral region reflecting the accumulated effect of metal absorbers (MgII, CIV and then SiIV) when probing higher redshifts.
We do not obtain a very good fit in the MgII region probably because of other unidentified contaminants (for instance in the ISM as studied by \cite{Vadai2017}). This would explain why we obtain from the auto-correlation a MgII bias larger that the one derived from the cross-correlation with quasars \cite{dMdB2019}. The signal is however four times smaller than the contributions of CIV and SiIV so we did not investigate this in more detail. On the contrary, we obtain good fits in the CIV and Si IV regions. We obtain a CIV bias consistent with result from the cross-correlation with quasars \cite{Blomqvist2018}. The increased bias we measure in the SiIV region is found compatible with studies performed for the Lyman-$\alpha$ 1D power spectrum analysis \cite{PalanqueDelabrouille2013,Ravoux2023}, and the ratio of SiIV to CIV bias is also found consistent with estimates from stacked absorption profiles from \cite{Pieri2014letter}. We have further demonstrated the validity of this approach by measuring the auto-correlation using pairs of quasar spectra from different spectrographs where only the astrophysical signal was measured (see Fig.~\ref{fig:metals-cf-2d-different-petals}). We measure a combined absorber bias $b^{eff}_{CIV}=-0.0191 \pm 0.0015$, with $\beta_{CIV}=0.5$. The correlated noise amplitude varies with the data set but can easily be separated from the astrophysical signal in a 2D correlation function.

The contamination from correlated noise and the CIV auto-correlation have been considered in the Lyman-$\alpha$ BAO analysis of the eBOSS data \cite{dMdB2020} albeit with a different functional form for the correlated noise given the different mapping between fibers in the focal plane and spectrographs. The amplitude of both contributions were fit as extra nuisance parameters. They measured for CIV a velocity bias $b_{\eta,CIV} = -0.0049 \pm 0.0026$ (see Table 6 of \cite{dMdB2020}). Given the assumed value of $\beta_{CIV} = 0.27$ and the growth rate $f \sim 0.94$ at $z_{CIV} \sim 1.6$, this gives a value of $b^{eff}_{CIV} = b_{\eta,CIV} \, f/\beta_{CIV} = -0.017 \pm 0.009$ that is uncertain but consistent with our measurement.

We note that other metals with wavelength close to the Lyman-$\alpha$ line also contaminate the forests. Those are Si II lines at 1190\AA, 1193\AA, and 1260\AA, and a Si III line at 1207\AA. Their contribution can be detected through their cross-correlation with Lyman-$\alpha$ and quasars. They produce extra peaks along the line of sight in the Lyman-$\alpha$ auto-correlation and the Lyman-$\alpha$ QSO cross-correlation. This signal is modeled and fit simultaneously with other parameters when analyzing the correlation functions (see e.g.~\cite{dMdB2020,KP6}). Those transitions, at shorter wavelengths, do not contribute to the measurements presented in this paper.

It is reassuring to note that for DESI the contamination from the instrumental effects is very small. The correlated noise signal is featureless at $r_\parallel >4$~\mpc\ and about 40 times smaller than the \lya\ signal at the BAO scale. We propose to use this two component contamination model (correlated noise and CIV auto-correlation) when analyzing the DESI Lyman-$\alpha$ auto-correlation function.

\begin{acknowledgments}

Special thanks to the DESI internal reviewer Julian Bautista for providing invaluable feedback on this manuscript.

AFR acknowledges financial support from the Spanish Ministry of Science and Innovation under the Ramon y Cajal program (RYC-2018-025210) and the PGC2021-123012NB-C41 project, and from the European Union's Horizon Europe research and innovation programme (COSMO-LYA, grant agreement 101044612). IFAE is partially funded by the CERCA program of the Generalitat de Catalunya.

 This material is based upon work supported by the U.S. Department of Energy (DOE), Office of Science, Office of High-Energy Physics, under Contract No. DE–AC02–05CH11231, and by the National Energy Research Scientific Computing Center, a DOE Office of Science User Facility under the same contract. Additional support for DESI was provided by the U.S. National Science Foundation (NSF), Division of Astronomical Sciences under Contract No. AST-0950945 to the NSF’s National Optical-Infrared Astronomy Research Laboratory; the Science and Technology Facilities Council of the United Kingdom; the Gordon and Betty Moore Foundation; the Heising-Simons Foundation; the French Alternative Energies and Atomic Energy Commission (CEA); the National Council of Science and Technology of Mexico (CONACYT); the Ministry of Science and Innovation of Spain (MICINN), and by the DESI Member Institutions: \url{https://www.desi.lbl.gov/collaborating-institutions}. Any opinions, findings, and conclusions or recommendations expressed in this material are those of the author(s) and do not necessarily reflect the views of the U. S. National Science Foundation, the U. S. Department of Energy, or any of the listed funding agencies.

The authors are honored to be permitted to conduct scientific research on Iolkam Du’ag (Kitt Peak), a mountain with particular significance to the Tohono O’odham Nation.

\end{acknowledgments}

\bibliography{biblio}{}
\bibliographystyle{JHEP}

\appendix


\section{Author Affiliations}
\label{sec:affiliations}

\noindent \hangindent=.5cm $^{1}${Lawrence Berkeley National Laboratory, 1 Cyclotron Road, Berkeley, CA 94720, USA}

\noindent \hangindent=.5cm $^{2}${IRFU, CEA, Universit\'{e} Paris-Saclay, F-91191 Gif-sur-Yvette, France}

\noindent \hangindent=.5cm $^{3}${Department of Physics and Astronomy, The University of Utah, 115 South 1400 East, Salt Lake City, UT 84112, USA}

\noindent \hangindent=.5cm $^{4}${Center for Cosmology and AstroParticle Physics, The Ohio State University, 191 West Woodruff Avenue, Columbus, OH 43210, USA}

\noindent \hangindent=.5cm $^{5}${NASA Einstein Fellow}

\noindent \hangindent=.5cm $^{6}${Institut de F'{i}sica d’Altes Energies (IFAE), The Barcelona Institute of Science and Technology, Campus UAB, 08193 Bellaterra Barcelona, Spain}

\noindent \hangindent=.5cm $^{7}${Institut de F\'{i}sica d’Altes Energies (IFAE), The Barcelona Institute of Science and Technology, Campus UAB, 08193 Bellaterra Barcelona, Spain}

\noindent \hangindent=.5cm $^{8}${Department of Physics \& Astronomy, University College London, Gower Street, London, WC1E 6BT, UK}

\noindent \hangindent=.5cm $^{9}${Departamento de F\'{i}sica, Universidad de Guanajuato - DCI, C.P. 37150, Leon, Guanajuato, M\'{e}xico}

\noindent \hangindent=.5cm $^{10}${Department of Astronomy, The Ohio State University, 4055 McPherson Laboratory, 140 W 18th Avenue, Columbus, OH 43210, USA}

\noindent \hangindent=.5cm $^{11}${Department of Physics, The Ohio State University, 191 West Woodruff Avenue, Columbus, OH 43210, USA}

\noindent \hangindent=.5cm $^{12}${The Ohio State University, Columbus, 43210 OH, USA}

\noindent \hangindent=.5cm $^{13}${Instituto de F\'{\i}sica, Universidad Nacional Aut\'{o}noma de M\'{e}xico,  Cd. de M\'{e}xico  C.P. 04510,  M\'{e}xico}

\noindent \hangindent=.5cm $^{14}${Aix Marseille Univ, CNRS, CNES, LAM, Marseille, France}

\noindent \hangindent=.5cm $^{15}${Departament de F\'isica, EEBE, Universitat Polit\`ecnica de Catalunya, c/Eduard Maristany 10, 08930 Barcelona, Spain}

\noindent \hangindent=.5cm $^{16}${Aix Marseille Univ, CNRS/IN2P3, CPPM, Marseille, France}

\noindent \hangindent=.5cm $^{17}${Universit\'{e} Clermont-Auvergne, CNRS, LPCA, 63000 Clermont-Ferrand, France}

\noindent \hangindent=.5cm $^{18}${Sorbonne Universit'{e}, CNRS/IN2P3, Laboratoire de Physique Nucl'{e}aire et de Hautes Energies (LPNHE), FR-75005 Paris, France}

\noindent \hangindent=.5cm $^{19}${IRFU, CEA, Universit'{e} Paris-Saclay, F-91191 Gif-sur-Yvette, France}

\noindent \hangindent=.5cm $^{20}${University Observatory, Faculty of Physics, Ludwig-Maximilians-Universit\"{a}t, Scheinerstr. 1, 81677 M\"{u}nchen, Germany}

\noindent \hangindent=.5cm $^{21}${Excellence Cluster ORIGINS, Boltzmannstrasse 2, D-85748 Garching, Germany}

\noindent \hangindent=.5cm $^{22}${Physics Dept., Boston University, 590 Commonwealth Avenue, Boston, MA 02215, USA}

\noindent \hangindent=.5cm $^{23}${Department of Physics and Astronomy, University of California, Irvine, 92697, USA}

\noindent \hangindent=.5cm $^{24}${SLAC National Accelerator Laboratory, Menlo Park, CA 94305, USA}

\noindent \hangindent=.5cm $^{25}${Kavli Institute for Particle Astrophysics and Cosmology, Stanford University, Menlo Park, CA 94305, USA}

\noindent \hangindent=.5cm $^{26}${Observatorio Astron\'omico, Universidad de los Andes, Cra. 1 No. 18A-10, Edificio H, CP 111711 Bogot\'a, Colombia}

\noindent \hangindent=.5cm $^{27}${Departamento de F\'isica, Universidad de los Andes, Cra. 1 No. 18A-10, Edificio Ip, CP 111711, Bogot\'a, Colombia}

\noindent \hangindent=.5cm $^{28}${Institut d'Estudis Espacials de Catalunya (IEEC), 08034 Barcelona, Spain}

\noindent \hangindent=.5cm $^{29}${Institute of Cosmology and Gravitation, University of Portsmouth, Dennis Sciama Building, Portsmouth, PO1 3FX, UK}

\noindent \hangindent=.5cm $^{30}${Institute of Space Sciences, ICE-CSIC, Campus UAB, Carrer de Can Magrans s/n, 08913 Bellaterra, Barcelona, Spain}

\noindent \hangindent=.5cm $^{31}${Consejo Nacional de Ciencia y Tecnolog\'{\i}a, Av. Insurgentes Sur 1582. Colonia Cr\'{e}dito Constructor, Del. Benito Ju\'{a}rez C.P. 03940, M\'{e}xico D.F. M\'{e}xico}

\noindent \hangindent=.5cm $^{32}${Fermi National Accelerator Laboratory, PO Box 500, Batavia, IL 60510, USA}

\noindent \hangindent=.5cm $^{33}${Department of Astrophysical Sciences, Princeton University, Princeton NJ 08544, USA}

\noindent \hangindent=.5cm $^{34}${NSF NOIRLab, 950 N. Cherry Ave., Tucson, AZ 85719, USA}

\noindent \hangindent=.5cm $^{35}${Department of Physics, Southern Methodist University, 3215 Daniel Avenue, Dallas, TX 75275, USA}

\noindent \hangindent=.5cm $^{36}${Sorbonne Universit\'{e}, CNRS/IN2P3, Laboratoire de Physique Nucl\'{e}aire et de Hautes Energies (LPNHE), FR-75005 Paris, France}

\noindent \hangindent=.5cm $^{37}${Departament de F\'{i}sica, Serra H\'{u}nter, Universitat Aut\`{o}noma de Barcelona, 08193 Bellaterra (Barcelona), Spain}

\noindent \hangindent=.5cm $^{38}${Instituci\'{o} Catalana de Recerca i Estudis Avan\c{c}ats, Passeig de Llu\'{\i}s Companys, 23, 08010 Barcelona, Spain}

\noindent \hangindent=.5cm $^{39}${Department of Astronomy, Tsinghua University, 30 Shuangqing Road, Haidian District, Beijing, China, 100190}

\noindent \hangindent=.5cm $^{40}${Department of Physics and Astronomy, Siena College, 515 Loudon Road, Loudonville, NY 12211, USA}

\noindent \hangindent=.5cm $^{41}${Department of Physics and Astronomy, University of Sussex, Brighton BN1 9QH, U.K}

\noindent \hangindent=.5cm $^{42}${Department of Physics \& Astronomy, University  of Wyoming, 1000 E. University, Dept.~3905, Laramie, WY 82071, USA}

\noindent \hangindent=.5cm $^{43}${National Astronomical Observatories, Chinese Academy of Sciences, A20 Datun Rd., Chaoyang District, Beijing, 100012, P.R. China}

\noindent \hangindent=.5cm $^{44}${Instituto Avanzado de Cosmolog\'{\i}a A.~C., San Marcos 11 - Atenas 202. Magdalena Contreras, 10720. Ciudad de M\'{e}xico, M\'{e}xico}

\noindent \hangindent=.5cm $^{45}${Department of Physics and Astronomy, University of Waterloo, 200 University Ave W, Waterloo, ON N2L 3G1, Canada}

\noindent \hangindent=.5cm $^{46}${Perimeter Institute for Theoretical Physics, 31 Caroline St. North, Waterloo, ON N2L 2Y5, Canada}

\noindent \hangindent=.5cm $^{47}${Waterloo Centre for Astrophysics, University of Waterloo, 200 University Ave W, Waterloo, ON N2L 3G1, Canada}

\noindent \hangindent=.5cm $^{48}${University of California, Berkeley, 110 Sproul Hall \#5800 Berkeley, CA 94720, USA}

\noindent \hangindent=.5cm $^{49}${Space Sciences Laboratory, University of California, Berkeley, 7 Gauss Way, Berkeley, CA  94720, USA}

\noindent \hangindent=.5cm $^{50}${Department of Physics, Kansas State University, 116 Cardwell Hall, Manhattan, KS 66506, USA}

\noindent \hangindent=.5cm $^{51}${Department of Physics and Astronomy, Sejong University, Seoul, 143-747, Korea}

\noindent \hangindent=.5cm $^{52}${CIEMAT, Avenida Complutense 40, E-28040 Madrid, Spain}

\noindent \hangindent=.5cm $^{53}${Department of Physics, University of Michigan, Ann Arbor, MI 48109, USA}

\noindent \hangindent=.5cm $^{54}${University of Michigan, Ann Arbor, MI 48109, USA}

\noindent \hangindent=.5cm $^{55}${Department of Physics \& Astronomy, Ohio University, Athens, OH 45701, USA}

\end{document}